\documentclass[ALICE,manyauthors]{cernphprep}

\usepackage[comma,square,numbers,sort&compress]{natbib}
\usepackage{hyperref}
\usepackage{lineno}
\usepackage{multirow}
\usepackage{soul}
\usepackage{rotating, graphicx}
\usepackage{lscape}

\usepackage{color}
\definecolor{orange}{rgb}{1,0.5,0}
\definecolor{darkgreen}{rgb}{0.0, 0.6, 0.22}

\newcommand{\lambdas}{$\Lambda$+$\overline{\Lambda}$}
\newcommand{\xis}{$\Xi^\pm$}
\newcommand{\omegas}{$\Omega^\pm$}

\newcommand{\pip}{\ensuremath{\pi^{+}}}
\newcommand{\pim}{\ensuremath{\pi^{-}}}

\newcommand{\pipm}{\ensuremath{\pi^{\pm}}}
\newcommand{\kapm}{\ensuremath{{\rm K}^{\pm}}}
\newcommand{\kanull}{\ensuremath{{\rm K}^{\rm 0}_{\rm S}}}

\newcommand{\vo}{\ensuremath{{\rm V}^{\rm 0}}}

\newcommand{\pp}{\ensuremath{\mathrm {p\kern-0.05em p}}}

\newcommand{\sqrtSnn}{\ensuremath{\sqrt{s_{\mathrm{NN}}}}}

\newcommand{\pt}{\ensuremath{p_{\mathrm{T}}}}
\newcommand{\vn}{\ensuremath{v_{\rm n}}}
\newcommand{\vnpt}{\ensuremath{v_{\rm n}(p_{\mathrm{T}})}}
\newcommand{\vtwopt}{\ensuremath{v_{\rm 2}(p_{\mathrm{T}})}}
\newcommand{\vthreept}{\ensuremath{v_{\rm 3}(p_{\mathrm{T}})}}
\newcommand{\vfourpt}{\ensuremath{v_{\rm 4}(p_{\mathrm{T}})}}

\newcommand{\vnmax}{\ensuremath{\pt{}\vert_{v_{\rm n}^{\mathrm{max}}}/n_{\rm q}}}
\newcommand{\vtwomax}{\ensuremath{\pt{}\vert_{v_2^{\mathrm{max}}}/n_{\rm q}}}
\newcommand{\vthreemax}{\ensuremath{\pt{}\vert_{v_3^{\mathrm{max}}}/n_{\rm q}}}

\newcommand{\dEdx}{\ensuremath{\mathrm{d}E/\mathrm{d}x}}

\newcommand{\MeVc}{\ensuremath{\mathrm{MeV}\kern-0.05em/\kern-0.02em c}}
\newcommand{\GeVc}{\ensuremath{\mathrm{GeV}\kern-0.05em/\kern-0.02em c}}
\newcommand{\GeVcSq}{\ensuremath{\mathrm{GeV}\kern-0.05em/\kern-0.02em c^2}}

\newcommand{\pbar}{\ensuremath{\rm\overline{p}}}
\newcommand{\ppbar}{\ensuremath{\rm p+\overline{p}}}
\newcommand{\jpsi}{\ensuremath{{\rm J}\kern-0.02em/\kern-0.05em\psi}}

\newcommand{\dpt}{\ensuremath{\delta \kern-0.15em p_{\mathrm{T}}}}

\newcommand{\rpn}{\ensuremath{\Psi_{\rm n}}}

\newcommand{\trento}{T\raisebox{-.5ex}{R}ENTo}
\usepackage{float}

\begin{document}

\begin{titlepage}
\PHyear{2018}
\PHnumber{103}      
\PHdate{10 May}  
%

\title{Anisotropic flow of identified particles in Pb--Pb collisions at $\mathbf{\sqrt{{\textit s}_{\rm NN}}}=5.02$~TeV}
\ShortTitle{Anisotropic flow of identified particles}   

\Collaboration{ALICE Collaboration\thanks{See Appendix~\ref{app:collab} for the list of collaboration members}}
\ShortAuthor{ALICE Collaboration} 


\begin{abstract}

The elliptic ($v_2$), triangular ($v_3$), and quadrangular ($v_4$) flow coefficients of $\pi^{\pm}$, ${\rm K}^{\pm}$, ${\rm p+\overline{p}}$, ${\Lambda+\overline{\Lambda}}$, ${\rm K}^{\rm 0}_{\rm S}$, and the $\phi$-meson are measured in Pb--Pb collisions at $\sqrt{s_{\rm NN}}=5.02$ TeV. Results obtained with the scalar product method are reported for the rapidity range $\vert y \vert <$ 0.5 as a function of transverse momentum, $p_\text{T}$, at different collision centrality intervals  between 0--70\%, including ultra-central (0--1\%) collisions for $\pi^{\pm}$, ${\rm K}^{\pm}$, and ${\rm p+\overline{p}}$. For $p_\text{T}~<~3$~GeV$\kern-0.05em/\kern-0.02em c$, the flow coefficients  exhibit a particle mass dependence. At intermediate transverse momenta ($3<p_\text{T}<$~8-10~GeV$\kern-0.05em/\kern-0.02em c$), particles show an approximate grouping  according to their type (i.e., mesons and baryons). The $\phi$-meson $v_2$, which tests both particle mass dependence and type scaling, follows ${\rm p+\overline{p}}$ $v_2$ at low $p_\text{T}$ and $\pi^{\pm}$ $v_2$ at intermediate $p_\text{T}$. The evolution of the shape of $v_{\rm n}(p_{\mathrm{T}})$ as a function of centrality  and harmonic number $n$ is studied for the various particle species. Flow coefficients of $\pi^{\pm}$, ${\rm K}^{\pm}$, and ${\rm p+\overline{p}}$ for $p_\text{T}<3$ GeV$\kern-0.05em/\kern-0.02em c$  are compared to iEBE-VISHNU and MUSIC hydrodynamical calculations coupled to a hadronic cascade model (UrQMD). The iEBE-VISHNU  calculations describe the results fairly well for $p_\text{T}~<~2.5$~GeV$\kern-0.05em/\kern-0.02em c$, while MUSIC calculations reproduce the measurements for $p_\text{T}~<~1$~GeV$\kern-0.05em/\kern-0.02em c$.  A comparison to $v_{\rm n}$ coefficients measured in Pb--Pb collisions at $\sqrt{s_{\rm NN}}$~=~2.76~TeV is also provided.
\end{abstract}

\end{titlepage}

\setcounter{page}{2}

\section{Introduction}

Ultra-relativistic heavy-ion collisions are used to study the properties of the quark-gluon plasma (QGP), a state of 
deconfined quarks and gluons expected at high temperatures or baryon densities \cite{qgp}. Measurements of anisotropies in 
particle azimuthal distributions relative to the collision symmetry planes at the Relativistic Heavy Ion Collider 
(RHIC)~\cite{brahmsqgp,phenixqgp,phobosqgp,fluid} and the Large Hadron Collider (LHC)~\cite{aliceqgp,atlasqgp,cmsqgp} have shown 
that the produced hot and dense matter behaves as a strongly-interacting QGP. Comparisons to predictions from hydrodynamic models 
indicate that the QGP has a shear viscosity to entropy density ratio $(\eta/s)$ close to the theoretical lower limit from the anti-de 
Sitter/conformal field theory (AdS/CFT) correspondence of $1/4\pi$ for $\hbar=k_{\rm B}=1$ \cite{kovtun}.

Azimuthal anisotropies in particle production relative to the collision symmetry planes, often referred to as anisotropic flow, arise from the 
asymmetry in the initial geometry of the collision combined with the initial inhomogeneities of the system's energy density \cite{olli}. Anisotropic flow 
depends on the equation of state and transport coefficients of the system, such as $\eta/s$ and bulk viscosity to entropy density 
ratio ($\zeta/s$). Its magnitude is quantified via the coefficients $v_{\rm n}$ in a Fourier decomposition of the particle 
azimuthal distribution~\cite{zhang} 
\begin{equation}
\label{eq:flowdud}
E\frac{\mathrm{d}^{3}N}{\mathrm{d}p^3} = \frac{1}{2\pi} \frac{\mathrm{d}^{2}N}{p_{\rm T}\mathrm{d} \pt{} \mathrm{d} y}(1+2\sum_{\rm n=1}^{\infty}v_{\rm n} \cos[{\rm n}(\varphi - \Psi_{\rm n})]),
\end{equation}
where $E$ is the energy, ${p}$ the momentum, \pt{} the transverse momentum, $\varphi$ the azimuthal angle, $\eta$ the pseudorapidity 
of the particle, and \rpn{} the $n$-th harmonic symmetry plane angle. The second order 
flow coefficient $v_2$, called \emph{elliptic flow}, is the largest contribution to the asymmetry of
non-central collisions because of the almond-like geometry of the overlap region between the colliding nuclei in the plane perpendicular to 
the beam direction. The third-order flow coefficient $v_3$, named \emph{triangular flow}, is generated by fluctuations in the initial 
distribution of nucleons and gluons in the overlap region \cite{fluc1,fluc2,fluc3,fluc4}. The fourth-order flow coefficient $v_4$, called 
\emph{quandrangular flow}, is generated both by initial geometry, fluctuations, and is in addition sensitive to the non-linear hydrodynamic 
response of the medium \cite{nl1,nl2}. It has been shown that higher-order flow coefficients are more sensitive to $\eta/s$ than 
$v_2$ \cite{sens1,sens2}. 

In addition to probing $\eta/s$ and $\zeta/s$, anisotropic flow constrains the initial spatial density (e.g.\ energy and entropy density), freeze-out 
conditions of the system, and particle production mechanisms in different \pt{} regions. Stronger constraints are achieved 
by studying anisotropic flow of identified particles. To guide interpretation of the results in the context of these processes, three kinematic 
`regions of interest' are defined in the \pt-differential $v_{\rm n}$ measurements, \vnpt{}. For $\pt \lesssim$ 3 \GeVc{}, anisotropic flow is a 
remnant of the collective dynamics during the hydrodynamic expansion of the system. The interplay between the isotropic expansion 
(radial flow) and anisotropic flow leads to a characteristic mass ordering of 
\vnpt{} \cite{massdep1,massdep2,massdep3,massdep4,massdep5,massdep7,massdep8,Abelev:2014pua, highham}, meaning that heavier 
particles have smaller \vnpt{}. At intermediate \pt{} (3 $\lesssim \pt{} \lesssim$ 8 \GeVc{}), the values of $v_{\rm n}$ for different particles tend to separate 
mesons and baryons \cite{Abelev:2007qg,Abelev:2007rw,scalingphenix,Adare:2012vq,highpt,Abelev:2014pua,highham}. The flow of baryons is 
larger than that of mesons in this \pt{} range, 
supporting the hypothesis of hadronization through quark coalescence \cite{dud2}, where it is assumed that the invariant spectrum of produced 
particles is proportional to the product of the spectra of their constituents \cite{coalescence,coalescence1}. However, the scaling only holds approximately 
at RHIC \cite{Adare:2012vq} and at the level of $\pm 20\%$ in Pb--Pb collisions at \sqrtSnn{} = 2.76 TeV \cite{Abelev:2014pua, highham}. This 
behaviour is also qualitatively consistent with a scenario in which particle production includes interactions of jet fragments 
with bulk matter~\cite{Werner:2012xh, Werner:2012ca}. For \pt{} $\gtrsim$ 8 \GeVc{}, anisotropic flow is generated when hard partons 
that propagate through the system lose energy via (multiple) scattering and gluon radiation \cite{gluonradiation,gluonradiation2}, resulting in 
$v_{\rm n}$ that remain non-zero up to very high \pt{} \cite{atlas,alicejet,Sirunyan:2017pan, Acharya:2018lmh}.

Anisotropic flow of identified particles is an important observable when studying the characteristics of the QGP. However, since particles can scatter 
and be regenerated in between the chemical and kinetic freeze-out of a collision (the \emph{hadronic phase}), information about the QGP phase 
imprinted in \vnpt{} can be altered by late-stage interactions and resonance decays, which can affect both \vn{} 
and $\langle \pt{} \rangle$ \cite{rescatter}, leading to a 
deviation in mass ordering in \vnpt{} at low \pt{} \cite{rescatter1}. The $\phi$-meson has been suggested as a particularly sensitive probe of the 
early collision phase as its production rate via regeneration in the hadronic phase is negligible \cite{starphi} and it is theorized to have a low 
hadronic cross section \cite{phics,phics1,phics2}, making it insensitive to the dissipative effects of the hadronic phase of the collision (although 
it should be noted that there is no consensus on the exact value of the cross section between the $\phi$-meson and nucleons in heavy-ion 
collisions \cite{phics3,phics4,phics5,phics6}). Recent experimental studies \cite{Abelev:2014pua,phiprodalice,hades} suggest that the $\phi$-meson may 
be more sensitive to the hadronic phase than anticipated. 

In this article, we present measurements of \pt-differential elliptic, triangular, and quadrangular flow coefficients 
of \pipm{}, \kapm{}, p+\pbar{}, \lambdas{}, \kanull{}, and the $\phi$-meson in Pb--Pb collisions at $\sqrt{s_{\rm NN}}$~=~5.02 TeV, extending greatly, and improving in precision upon, the previous measurements of identified particle
 $v_{\rm n}$ in Pb--Pb collisions at $\sqrt{s_{\rm NN}}$~=~2.76 TeV as carried out by ALICE \cite{Abelev:2014pua,highham,highpt}. The 
results are reported for a wide range of particle transverse momenta within the rapidity range $\vert y \vert <$ 0.5 at different collision 
centralities between 0--70\% range. To isolate the fraction of anisotropic flow that is generated by initial-state fluctuations rather than 
geometry, the flow coefficients are also studied in ultra-central collisions (0--1\% collision centrality). Centrality estimates the degree of 
overlap between the two colliding nuclei and is expressed as percentiles of the inelastic hadronic cross section, with low percentage values 
corresponding to head-on collisions. The measurements are performed using the scalar product method~\cite{Adler:2002pu,Voloshin:2008dg,Luzum:2012da} 
with a (pseudo-)rapidity gap of $|\Delta\eta|>2.0$ between the identified particles under study and the charged reference particles. The flow 
coefficients are measured separately for particles and anti-particles and are found to be compatible within the statistical uncertainties for 
most \pt{} and centrality intervals. Any residual differences are included in the systematic uncertainties, and $v_{\rm n}$ denotes the 
average between results for particles and anti-particles.

This paper is organized as follows. Analysis details, particle identification, reconstruction methods, and flow measurement techniques 
are outlined in Sec.~\ref{sec:analysis}. The evaluation of systematic uncertainties is discussed in Sec.~\ref{sec:systematics}. The flow 
coefficients of \pipm{}, \kapm{}, p+\pbar{} ($v_2$, $v_3$, and $v_4$), \lambdas{}, \kanull{} ($v_2$ and $v_3$), and the $\phi$-meson ($v_2$) 
are reported and compared to model calculations in Sec.~\ref{sec:results}. Finally, the results are summarized in Sec.~\ref{sec:summary}.

\section{Experimental setup and data analysis}
\label{sec:analysis}

ALICE~\cite{Carminati:2004fp, Alessandro:2006yt, review} is a dedicated heavy-ion experiment at the LHC optimized to study the properties of 
strongly interacting matter produced in heavy-ion collisions. A full overview of the detector layout and its performance can be found 
in \cite{review,performance}. The main subsystems used in this analysis are the Inner Tracking System (ITS)~\cite{its}, Time Projection 
Chamber (TPC)~\cite{tpc}, Time Of Flight detector (TOF)~\cite{tof}, and V0 \cite{vzero}. The ITS, TPC, and TOF detectors cover full 
azimuth within pseudorapidity range 
$\vert \eta \vert <$ 0.9 and lie within a homogeneous magnetic field of up to 0.5 T. The ITS consists of six layers of silicon 
detectors used for tracking and vertex reconstruction. The TPC is the main tracking detector and is also used to identify particles 
via specific ionization energy loss, $\mathrm{d}E/\mathrm{d}x$. The TOF in conjunction with the timing information from the T0 detector~\cite{t0} 
provide particle identification based on flight time. The T0 is made up of two arrays of Cherenkov counters T0C and T0A, located 
at -3.3 $< \eta <$ -3.0 and 4.5 $< \eta <$ 4.9, respectively. Two scintillator arrays (V0), which cover the pseudorapidity 
ranges $-3.7<\eta<-1.7$ (V0C) and $2.8<\eta<5.1$ (V0A), are used for triggering, event selection, and the determination of 
centrality~\cite{centrality} and $\textbf{Q}_{\rm n}$-vectors (see Sec.~\ref{sec:flow}). Both V0 detectors are segmented in four rings in the radial 
direction with each ring divided into eight sectors in the azimuthal direction. In addition, two tungsten-quartz neutron Zero Degree Calorimeters (ZDCs), installed 112.5 meters from the interaction point on each side, are used for event selection. 

\subsection{Event and track selection}\label{sec:evtrack}

The data sample recorded by ALICE during the 2015 LHC Pb--Pb run at \sqrtSnn~=~5.02~TeV is used for this analysis. The minimum-bias 
trigger requires signals in both V0A and V0C detectors. An offline event selection is applied to remove beam-induced background (i.e.\ 
beam-gas events) and pileup events. The former is rejected utilizing the V0 and ZDC timing information. The 
remaining contribution of such interactions is found to be smaller than 0.02\%~\cite{performance}. Pileup events, which constitute 
about 0.25\% of the recorded sample, are removed by 
comparing multiplicity estimates from the V0 detector to those of tracking detectors at mid-rapidity, exploiting the difference in readout times 
between the systems. The fraction of pileup events left after applying the dedicated pileup removal criteria is found to be negligible. The primary vertex position is determined from tracks reconstructed in the ITS and TPC as described in 
Ref.~\cite{performance}. Only events with a primary vertex position within $\pm 10$~cm from the nominal interaction point along the beam 
direction are used in the analysis. Approximately $67 \times 10^6$ Pb--Pb events in the 0--70\% centrality interval pass 
these selection criteria. Centrality is estimated from the energy deposition measured in the V0 detector~\cite{centrality}.

Charged-particle tracks, used to measure the $v_{\rm n}$ of \pipm{}, \kapm{}, p+\pbar{} and the $\phi$-meson, are reconstructed using the ITS and TPC within $|\eta|<0.8$ and $0.5<\pt<16.0$ GeV/$c$ with a track-momentum 
resolution better than 4\% for the considered range~\cite{performance}. Additional quality criteria are used 
to reduce the contamination from secondary charged particles (i.e.,\ particles originating from weak decays, $\gamma$-conversions, and secondary 
hadronic interactions in the detector material) and fake tracks (random associations of space points). Only tracks with at least 70 
space points, out of a maximum of 159, with a $\chi^2$ per degree-of-freedom for the track fit 
lower than 2, are accepted. Moreover, each track is required to cross at least 70 TPC pad rows and to 
be reconstructed from at least 80\% of the number of expected TPC space points, in addition to having at least one hit in the two innermost layers of the 
ITS. Furthermore, tracks with a distance of closest approach (DCA) to the reconstructed event vertex smaller than 2~cm in the longitudinal 
direction ($z$) and (0.0105 + 0.0350 (\pt{}~$c$/GeV)$^{-1.1}$)~cm in the transverse plane ($xy$) are selected. Relevant selection criteria for tracks used 
for the reconstruction of \kanull{} and \lambdas{} are given in Sec.~\ref{sec:v0s}.

\subsection{Identification of \pipm{}, \kapm{} and p+\pbar{}}
\label{sec:pid}

Particle identification is performed using the specific ionization energy loss, \dEdx{}, measured in the TPC and the time of flight 
obtained from the TOF. A truncated-mean procedure is used to estimate the \dEdx{} (where the 40\% highest-charge clusters are discarded), 
which yields a \dEdx{} resolution around 6.5\% in the 0--5\% centrality class~\cite{performance}. At least 70 clusters are used for the \dEdx{} 
estimation. The TOF measures the time that a particle needs to travel from the primary vertex to the detector itself with a time resolution of $\approx$ 80 
ps \cite{performance}. The start time for the TOF measurement is provided by the T0 detector or from a combinatorial algorithm which uses 
the particle arrival times at the TOF detector itself~\cite{performance,tof}.

Expressing the difference between the expected \dEdx{} and the time of flight for \pipm{}, \kapm{} and p+\pbar{}, and the measured signals in both TPC and TOF, in units of the standard deviations from the most probable value for both detectors (n$\sigma_{\rm TPC},~$n$\sigma_{\rm TOF}$), and applying a selection on the number of accepted n$\sigma$, allows for particle identification on a track-by-track basis. The TPC \dEdx{} of different particle species are separated by at least 4$\sigma$ for \pt{} $<$ 0.7 \GeVc{}, while in the relativistic rise region of the \dEdx{} ($\pt>$ 2 \GeVc{}) particle identification is still possible but only on a statistical basis \cite{performance}. The TOF detector provides 3$\sigma$ separation between \pipm{} and \kapm{} for \pt{} $<$ 2.5 \GeVc{}, and between \kapm{} and p+\pbar{} for \pt{} $<$ 4 \GeVc{} \cite{performance}.

The information from the TPC and TOF is combined using a quadratic sum 
${\rm n}\sigma_{\rm PID} = \sqrt{{\rm n}\sigma_{\rm TPC}^2 + {\rm n}\sigma_{\rm TOF}^2}$ for 0.5 $<$ \pt{} $\leq$ 4 \GeVc{}. Particles 
are selected by requiring ${\rm n}\sigma_{\rm PID} < 3$ for each species. The smallest ${\rm n}\sigma_{\rm PID}$ is used to assign the identity 
when the selection criterion is fulfilled by more than one species. When measuring p+\pbar{} \vnpt{}, only \pbar{} are considered for 
\pt{} $<$ 2 \GeVc{} to exclude secondary protons from detector material. At high transverse momenta 
(\pt{} $>$ 4~\GeVc{}), \kapm{} cannot reliably be identified. Identification of \pipm{} and p+\pbar{} for \pt{} $> 4$ \GeVc{} is done utilizing the 
TPC \dEdx{} signal only. Pions (protons) are selected from the upper (lower) part of the expected pion (proton) \dEdx{} distribution. For example, 
proton selection typically varies in the range from 0 to $-3\sigma_{\rm TPC}$ or from $-1.5\sigma_{\rm TPC}$ to $-4.5\sigma_{\rm TPC}$ depending on the momentum.

Secondary contamination from weak decays, studied using the procedure outlined in \cite{Abelev:2013vea}, decreases from about 30\% to 5\% for p+\pbar{} in the \pt{} range  0.7-4.0~GeV/$c$ and from 
about 5\% to 0.5\% for \pipm{} in the \pt{} range 0.5-4.0~GeV/$c$, while it is negligible for \kapm{}. The $v_{\rm n}$ coefficients are not corrected for these contaminations; their effect on $v_{\rm n}$ is at maximum $\approx$ 8\%, for \ppbar{} $v_2$ at \pt{} $<$ 1 \GeVc{} for central collisions, and negligible for \kapm{}, \pipm{} $v_{\rm n}$. The contamination from other particle species is below 3\% and 20\% at \pt{} $>$ 4.0 \GeVc{} for \pipm{} and p+\pbar{}, 
respectively, and contamination from fake tracks is negligible. The $v_{\rm n}$ results are reported for 
$0.5<\pt<16.0 (12.0, 6.0)$~\GeVc{} for \pipm{} $v_2$ ($v_3$, $v_4$), $0.7<\pt<16.0 (12.0, 6.0)$~\GeVc{} for p+\pbar{} $v_2$ 
($v_3$, $v_4$), and $0.5<\pt<4.0$~\GeVc{} for \kapm{} $v_{\rm n}$, all within $\vert y \vert <$ 0.5.

\subsection{Reconstruction of \kanull{} and $\Lambda+\overline{\Lambda}$}
\label{sec:v0s}

The \kanull{} and \lambdas{} are reconstructed in the \kanull{} $\rightarrow$ \pip{} + \pim{} and 
$\Lambda{}$ $\rightarrow$ p + \pim{} ($\overline{\Lambda} \rightarrow \pbar{} + \pip{}$) channels with branching ratios of 69.2\% \cite{pdg2016} and 63.9\% \cite{pdg2016} respectively. Reconstruction of \kanull{} and \lambdas{} is based on identifying secondary vertices from which two oppositely-charged 
particles originate, called V$^0$s. Topological selection criteria pertaining to the shape of the \vo{} decay can be imposed, as 
well as requirements on the species identity of the decay products (called \emph{daughter particles}).

The \vo{} candidates are selected to have an invariant mass between 0.4 and 0.6 GeV/$c^2$ and 1.07 and 1.17 GeV/$c^2$ for \kanull{} 
and \lambdas{}, respectively. The invariant mass of the \vo{} is calculated based on the assumption that 
the daughter particles are either a \pip\pim{} pair, or a p\pim{} (\pbar\pip{}) pair. The daughter particles have been 
identified over the entire $\pt$ range using the TPC following the n$\sigma$ approach detailed in 
Sec.~\ref{sec:pid} ($\vert {\rm n}\sigma_{\rm TPC} \vert < 3$). The daughter 
tracks were reconstructed within $|\eta|<0.8$, while the criteria on the number of TPC space points, the $\chi^2$ per TPC space point per 
degree-of-freedom, the number of crossed TPC pad rows, and the percentage of the expected TPC space points used to reconstruct a track 
are identical to those applied for primary particles. In addition, the minimum DCA of daughter tracks to the primary vertex is 0.1~cm. 
Furthermore, the maximum DCA of daughter tracks to the secondary vertex is 0.5~cm to ensure that they are products of the same decay.

To reject secondary vertices arising from decays into more than two particles, the cosine of the pointing angle $\theta_p$ is required to 
be larger than 0.998. This angle is defined as the angle between the momentum-vector of the \vo{} assessed at its decay position and the line connecting the \vo{} decay vertex to the 
primary vertex and has to be close to 0 as a result of momentum conservation. In addition, the \vo{} candidates are only accepted when they 
are produced at a distance between 5 and 100~cm from the nominal primary vertex in the radial direction. The lower value is chosen to avoid any bias from the efficiency loss when secondary tracks are being wrongly matched to clusters in the first layer of the ITS. To assess the systematic uncertainty related to contaminations from \lambdas{} and 
electron--positron pairs coming from $\gamma$-conversions to the \kanull{} sample, a selection in the Armenteros-Podolanski 
variables~\cite{armpod} is applied for the \kanull{} candidates, rejecting ones with $q \leq \vert \alpha \vert/5$. Here $q$ is the momentum 
projection of the positively charged daughter track in the plane perpendicular to the \vo{} momentum and 
$\alpha$ = ($p_{\rm L}^+ - p_{\rm L}^-)/(p_{\rm L}^+ + p_{\rm L}^-$), with $p_{\rm L}^{\pm}$ the projection of the positive or negative daughter tracks' momentum 
onto the momentum of the \vo{}. 

To obtain the \pt{}-differential yield of \kanull{} and \lambdas{} (which, together with background yields, are used for the \vn{} extraction cf. Eq. 4), invariant mass distributions at various \pt{} intervals are parametrized as a 
sum of a Gaussian distribution and a second-order polynomial function. The latter is introduced to account for residual contaminations 
(\emph{background yield}) that are present in the \kanull{} and \lambdas{} signals after the topological and daughter track 
selections. The \kanull{} and \lambdas{} yields are extracted by integration of the Gaussian distribution. Obtained yields have not been corrected for feed-down from higher mass baryons (\xis{}, \omegas{}) as earlier studies have shown that these have a negligible effect on the measured $v_{\rm n}$ \cite{Abelev:2014pua}. The \vnpt{} results are reported within $\vert y \vert <$ 0.5 and
$0.5<\pt<10$~\GeVc{} for \kanull{} and $0.8<\pt<10$~\GeVc{} for \lambdas{}.

\subsection{Reconstruction of $\phi$-mesons}
\label{sec:phi}

The $\phi$-meson is reconstructed in the $\phi \rightarrow$ K$^+$+K$^-$ channel with a branching ratio of 48.9\% \cite{pdg2016}. Its 
reconstruction proceeds by first identifying all primary \kapm{} tracks in an event, following the procedure for primary charged \kapm{} 
outlined in Sec.~\ref{sec:pid}. The \kapm{} identification criterion ${\rm n}\sigma_{\rm PID} < 3$ is chosen as it improves the significance 
of the $\phi$-meson yield, while retaining a sufficient reconstruction efficiency. The vector sums of all 
possible \kapm{} pairs are called $\phi$-meson \emph{candidates}, the yield of which is obtained as function of invariant mass 
M$_{\rm K^+ K^-}$ in various \pt{} intervals. The \pt{}-differential $\phi$-meson yield is obtained by first subtracting a \emph{background} yield 
from the candidate yield. This background yield is estimated using an event-mixing technique \cite{em}, in which \kapm{} from different 
collisions are paired into background tracks, and is 
normalized to the candidate yield for 1.04 $<$ M$_{\rm K^+ K^-} <$ 1.09 \GeVcSq{}. Collisions with similar characteristics (vertex position, 
centrality) are used for this mixing. To obtain 
the \pt{}-differential yield of $\phi$-mesons, the invariant mass distributions of the candidate yield is, after the aforementioned 
subtraction, parametrized as a sum 
of a Breit-Wigner distribution and a second-order polynomial function, the latter introduced to account for residual contaminations. The $\phi$-meson 
yields are extracted by integration of the Breit-Wigner distribution and, together with background yields, used for the \vn{} 
extraction (see Eq.~\ref{eq:invmassfit}). The \vtwopt{} results are reported for $0.9<\pt<6.5$~\GeVc{} within $\vert y \vert <$ 0.5.

\subsection{Flow analysis techniques}
\label{sec:flow}

The flow coefficients $v_{\rm n}$ are measured using the scalar product method \cite{Adler:2002pu,Voloshin:2008dg,Luzum:2012da}, written as
\begin{equation}
    v_{\rm n}\{{\rm SP}\} = \langle \langle {\bf u}_{\rm n, k} {\bf Q}_{\rm n}^{*} \rangle \rangle \Bigg/ \sqrt{ \frac{\langle {\bf Q}_{\rm n}  {\bf Q}_{\rm n}^{\rm A *} \rangle \langle  {\bf Q}_{\rm n} {\bf Q}_{\rm n}^{\rm B *} \rangle} { \langle  {\bf Q}_{\rm n}^{\rm A} {\bf Q}_{\rm n}^{\rm B *} \rangle } },
    \label{eq:mth_sp}
\end{equation}
where ${\bf u}_{\rm n, k} =\exp(in\varphi_k)$ is the unit flow vector of the particle of interest $k$ with azimuthal angle $\varphi_k$, ${\bf Q}_{\rm n}$ 
is the event flow vector, and $n$ is the harmonic number. Brackets $\langle \cdots \rangle$ denote an average over all events, the double 
brackets $\langle \langle \cdots \rangle \rangle$ an average over all particles in all events, and $^*$ the complex conjugate.

The vector ${\bf Q}_{\rm n}$ is calculated from the azimuthal distribution of the energy deposition measured in the V0A. Its $x$ and $y$ components 
are given by
\begin{equation}
\label{eq:Qcomp}  
Q_{\rm n,x} = \sum_j w_j \cos({\rm n} \varphi_j), \; Q_{\rm n,y} = \sum_j w_j \sin({\rm n} \varphi_j),
\end{equation}
where the sum runs over the 32 channels $j$ of the V0A detector, $\varphi_j$ is the azimuthal angle of channel $j$ defined by the geometric 
center, and $w_j$ is the amplitude measured in channel $j$. The vectors ${\bf Q}_{\rm n}^{\rm A}$ and ${\bf Q}_{\rm n}^{\rm B}$ are determined from 
the azimuthal distribution of the energy deposition measured in the V0C and the azimuthal distribution of the tracks reconstructed in 
the ITS and TPC, respectively. The amplitude measured in each channel of the V0C (32 channels as for the V0A) is used as weight in the 
case of ${\bf Q}_{\rm n}^{\rm A}$, while unity weights are applied for ${\bf Q}_{\rm n}^{\rm B}$. Tracks used for ${\bf Q}_{\rm n}^{\rm B}$ are selected following the procedure for primary charged tracks outlined in Sec.~\ref{sec:evtrack} for $0.2~<~\pt{}~<~5.0$ \GeVc{}. In order to account for a non-uniform detector 
response, the components of the ${\bf Q}_{\rm n}$ and ${\bf Q}_{\rm n}^{\rm A}$ vectors are recalibrated using a recentering procedure 
(i.e.\ subtraction of the ${\bf Q}_{\rm n}$-vector averaged over many events from the ${\bf Q}_{\rm n}$-vector of each event)~\cite{twist}. The large gap in pseudorapidity between \textbf{u}$_{\rm n, k}$ and ${\bf Q}_{\rm n}$ ($|\Delta\eta|>2.0$) greatly suppresses short-range 
correlations unrelated to the azimuthal asymmetry in the initial geometry, commonly referred to as `non-flow'. These correlations largely 
come from the inter-jet correlations and resonance decays.

\begin{figure}
    \begin{center}
    \includegraphics[width=.5\textwidth]{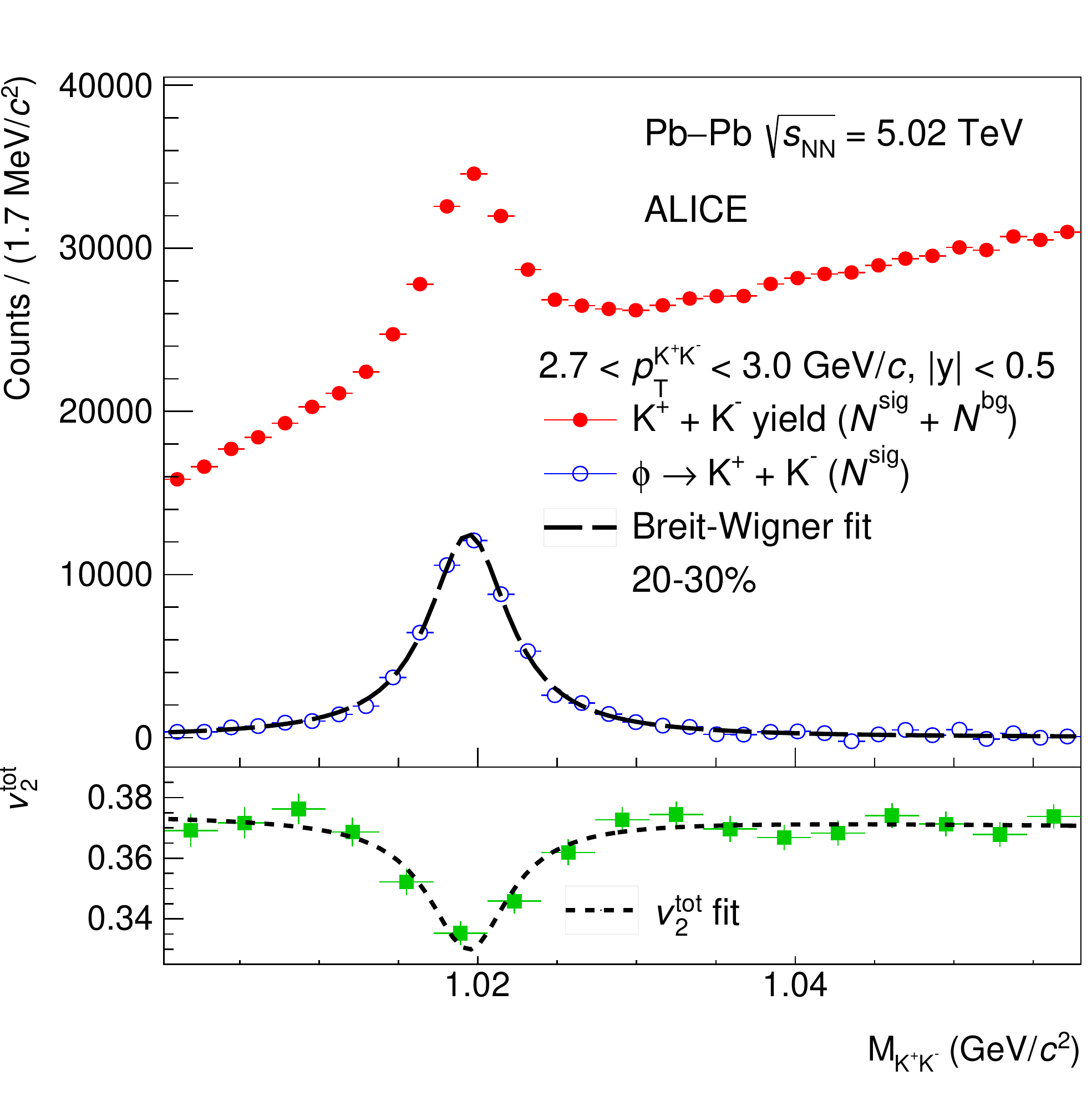}
    \caption{(Colour online) Illustration of reconstruction and $v_2$ measurement for the $\phi$-meson. The reconstruction of the $\phi$-meson and extraction of $N^{\rm sig}$ and $N^{\rm bg}$ are shown in the upper panel. A fit of Eq.~\ref{eq:invmassfit} to data is presented in the lower panel.}
        \label{fig:phi_method}
    \end{center}
\end{figure}

The $v_{\rm n}$ of the \kanull{}, \lambdas{}, and $\phi$-meson cannot directly be measured using Eq.~\ref{eq:mth_sp} as \kanull{}, \lambdas{} and 
the $\phi$-meson cannot be identified on a particle-by-particle basis. Therefore, the $v_{n}^{\rm{tot}}$ of \vo{}s and $\phi$-meson 
candidates is measured as function of both invariant mass, M$_{\rm d^+ d^-}$, and candidate \pt{}. This $v_{n}^{\rm{tot}}$ can be 
written \cite{borghini} as the weighted sum of \vnpt{} of the particle of interest, $v_{n}^{\rm{sig}}$, and that of background tracks, 
$v_{n}^{\rm{bg}}({\rm{M}_{\rm{d}^+\rm{d}^-}})$, as  
\begin{equation}
\label{eq:invmassfit}
    v_{\rm n}^{\rm{tot}} ({\rm{M}_{\rm{d}^+\rm{d}^-}}) = v_{\rm n}^{\rm{sig}} \frac{N^{\rm{sig}}}{N^{\rm{sig}}+ N^{\rm{bg}}}({\rm{M}_{\rm{d}^+\rm{d}^-}}) + v_{\rm n}^{\rm{bg}}({\rm{M}_{\rm{d}^+\rm{d}^-}}) \frac{N^{\rm{bg}}}{N^{\rm{sig}} + N^{\rm{bg}}}({\rm{M}_{\rm{d}^+\rm{d}^-}}),
\end{equation}
where \emph{signal} and \emph{background} yields $N^{\rm sig}$ and $N^{\rm bg}$ are obtained for each \pt{} interval from the \kanull{}, \lambdas{} and 
$\phi$-meson reconstruction procedures outlined in Secs~\ref{sec:v0s} and~\ref{sec:phi}. The formalism of Eq.~\ref{eq:mth_sp} is used 
to measure $v_{\rm n}^{\rm{tot}} (\rm{M}_{\rm{d}^+\rm{d}^-})$, $v_{\rm n}^{\rm{sig}}$ is obtained by parametrizing 
$v_{\rm n}^{\rm{bg}}({\rm{M}_{\rm{d}^+\rm{d}^-}})$ as a second-order polynomial function and fitting Eq.~\ref{eq:invmassfit} to the 
data. Figure~\ref{fig:phi_method} illustrates this procedure for the $\phi$-meson, showing the invariant mass spectrum of the $\phi$-meson in the 
upper panel, and a fit of Eq.~\ref{eq:invmassfit} to $v_{2}^{\rm{tot}} ({\rm{M}_{\rm{d}^+\rm{d}^-}})$ data in the lower panel.  

\section{Systematic uncertainties}
\label{sec:systematics}

The systematic uncertainties on $v_{\rm n}$ fall into the following categories: those arising from event selection, those arising from 
charged particle tracking, uncertainties in particle identification, uncertainties in \vo{} finding, and those coming from the extraction of \vnpt{}.

For $\pt \leq 4$~\GeVc{}, a \pt{}-dependent systematic uncertainty is assigned to $v_2$, $v_3$, and $v_4$  
of \pipm{}, \kapm{}, p+\pbar{}, \lambdas{}, \kanull{} and the $\phi$-meson. Per measured point, the difference between the nominal measurement 
and a variation on the nominal measurement is calculated. If this difference between the nominal data point and the systematic variation is 
significant (where significance is evaluated based on the recommendations in \cite{barlow}), it is considered to be a systematic uncertainty. When 
various checks are performed to quantify the effect of one systematic uncertainty (e.g. using three different centrality estimators to estimate the 
uncertainty in centrality determination), the maximum significant deviation that is found between the nominal measurement and the systematic 
variations is assigned as a systematic uncertainty. For each particle species, a \pt{}-independent average uncertainty is reported for 
$\pt{} > 4$ \GeVc{} in order to suppress sensitivity to statistical fluctuations.  The uncertainty is obtained by fitting a zeroth-order polynomial 
to the significant \pt{}-dependent relative uncertainties.

The systematic uncertainties are evaluated (if applicable) for each particle species, \vnpt{} and centrality 
intervals. A quadratic sum of the systematic uncertainties from the independent sources is reported as final systematic uncertainty on the 
measurements. An overview of the magnitude of the relative systematic uncertainties per particle species is given in 
Tabs.~\ref{tab:systematics1}, \ref{tab:systematics2}, and \ref{tab:systematics3} for $v_2$, $v_3$, and $v_4$, respectively.

 \begin{table}[t]
    \begin{center}
        \begin{tabular*}{160mm}{@{\extracolsep{\fill}}lcccccc}
            Error source & \pipm{} & \kapm{} & p+\pbar{}  & \kanull{} & \lambdas{} & $\phi$\\
            \hline\hline
            Vertex position & 0--1\% & 0--1\% & 0--2\% & 0--2\%& 0--4\%&  1--6\% \\
            1\% wide centrality intervals & 0--3\% & 0--4\% & 0--4\% & & & \\
            Centrality estimator & 0--3\% & 0--2\% & 0--3\% & 0--4\%& 0--5\%& 1--5\%  \\
            Magnetic field polarity & 0--2\% & 0--1\% & 0--2\% & 0--3\%& 0--3\%& 1--3\%  \\
            Interaction rate & 0--2\% & 0--1\% & 1--2\% & \emph{negl} & \emph{negl} & \emph{negl}   \\
            Pileup rejection & 0--1\% & 0--1\% & 0--2\% & 0--1\%&0--2\% & 0--1\%\\
            \hline
            Tracking mode & 0--4\% & 0--8\% &  0--10\% & & & 0--5\%\\
            Number of TPC space points & 0--2\% & 0--2\% & 0--2\% &0--4\% &0--2\% &  \emph{negl} \\
            Track quality & 0--3\% & 0--2\% & 0--3\% & 0--4\%& 0--3\% & \emph{negl} \\
            \hline
            Particle identification purity & 0--5\% & 0--7\% & 0--5\% & 0--3\%& 0--8\%& 0--6\% \\
            Number of TPC clusters used for \dEdx{} & 0--6\% & 0--5\% & 0--5\% & 0--5\%& 0\% & \emph{negl} \\
            Exclusive particle identification & 0--2\% & 0--3\% & 0--3\% & & & \\
            \hline 
            Decay vertex (radial position) & & & &0--10\% & 0--11\%&  \\
            Armenteros-Podolanski variables & & & &0--2\% & &  \\
            DCA decay products to primary vertex & & & &0--3\% & 0--5\%&   \\
            DCA between decay products & & & &0--2\% & 0--7\%&   \\
            Pointing angle $\cos\theta_{\rm{p}}$ & & & &0--4\% &0--9\% &   \\
            Minimum \pt{} of daughter tracks & & & &0--4\% &0--5\% & \\
            Peak shape & & & & \emph{negl}& \emph{negl} & \emph{negl}\\
            Residual background in yield & & & &\emph{negl} &\emph{negl} & \emph{negl}\\
            Event mixing & & & & & &1--3\% \\
            \hline 
            Positive and negative rapidities & 0--3\% & 0--2\% & 0--2\% &0--4\% &0--7\% & \emph{negl} \\
            Opposite charges & 0--2\% & 0--2\% & 0--2\% & & & \\
            Channel removal from V0A & 0--5\% & 0--5\% & 0--8\% & 0--3\%& 0--5\%& 0--4\% \\
            $v_{\rm n}$ from V0A or V0C & 0--2\% & 0--2\% & 0--2\% & \emph{negl} & \emph{negl}& \emph{negl}\\
            $v_{\rm n}^{\rm bg}$ parametrization & & & &\emph{negl} &\emph{negl}  &\emph{negl} \\
            $v_{n}^{\rm{tot}}$ fit ranges & & & & 0-1\% & 0-2\% & 0-1\% \\
            \hline
   \end{tabular*}
        \caption{\label{tab:systematics1} Summary of systematic uncertainties for the $v_2$ of \pipm{}, \kapm{}, p+\pbar{}, \lambdas{}, \kanull{}, and the $\phi$-meson. The uncertainties depend on \pt{} and centrality range; minimum and maximum values are listed here. Empty fields indicate that a given check does not apply to the particle of interest. If an uncertainty has been tested but cannot be resolved within statistical precision, the field is marked \emph{negl} for negligible. Horizontal lines are used to separate the different categories of systematic uncertainties as explained in Sec.~\ref{sec:systematics}.}
    \end{center}
\end{table}
\begin{table}[htbp]
    \begin{center}
        \begin{tabular*}{140mm}{@{\extracolsep{\fill}}lccccc}
            Error source & \pipm{} & \kapm{} & p+\pbar{} & \kanull{} & \lambdas{} \\
            \hline\hline
            Vertex position & 0--2\% & 0--1\% & 0--2\% & 0--3\%& 0--9\% \\
            1\% wide centrality intervals & 0--2\% & 0--2\% & 0--2\% & &  \\
            Centrality estimator & 0--2\% & 0--2\% & 0--2\% & 0--4\%&  0--9\% \\
            Magnetic field polarity & 0--2\% & 0--1\% & 0--3\% & 0--3\% & 0--3\%\\
            Interaction rate & 1--2\% & 1--2\% & 1--3\% & \emph{negl} &  \emph{negl}   \\
            Pileup rejection & 0--2\% & 0--1\% & 0--3\% & 0--1\%& 0--2\% \\
            \hline
            Tracking mode & 0--3\% & 1--5\% &  0--10\% & &  \\
            Number of TPC space points & 0--1\% & 0--2\% & 0--5\% &0--3\% & 0--6\% \\
            Track quality & 1--3\% & 1--2\% & 1--3\% & 0--3\%& 0--6\% \\
            \hline
            Particle identification purity & 0--4\% & 1--3\% & 0--10\% & 0--4\%& 0--4\% \\
            Number of TPC clusters used for \dEdx{} & 0--5\% & 0--5\% & 0--5\% & &  \\
            Exclusive particle identification & 0--1\% & 0--2\% & 0--1\% & &  \\
            \hline 
            Decay vertex (radial position) & & & &0--9\% &  0--11\% \\
            Armenteros-Podolanski variables & & & &0--4\% &   \\
            DCA decay products to primary vertex & & & &0--3\% & 0--5\%   \\
            DCA between decay products & & & &0--5\% & 0--8\%   \\
            Pointing angle $\cos\theta_{\rm{p}}$ & & & &0--5\% & 0--1\%  \\
            Minimum \pt{} of daughter tracks & & & &0--4\% & \emph{negl}  \\
            Peak shape & & & & \emph{negl}& \emph{negl}\\
            Residual background in yield & & & &\emph{negl} & \emph{negl} \\
            \hline 
            Positive and negative rapidities & 0--2\% & 0--1\% & 0--3\% &0--5\% & 0--4\% \\
            Opposite charges & 0--2\% & 0--2\% & 0--2\% & &  \\
            $v_{\rm n}$ from V0A or V0C & 0--2\% & 0--1\% & 0--2\% & 0--4\% & 0--3\% \\
            Channel removal from V0A & 0--8\% & 1--8\% & 1--8\% &0--4\% &  0--5\%\\
            $v_{\rm n}^{\rm bg}$ parametrization & & & &\emph{negl} &\emph{negl}  \\
            $v_{n}^{\rm{tot}}$ fit ranges & & & & 0-2\% & 0-2\%  \\
            \hline
   \end{tabular*}
   \caption{\label{tab:systematics2} Summary of systematic uncertainties for the $v_3$ of \pipm{}, \kapm{}, p+\pbar{}, \lambdas{}, and \kanull{}. The uncertainties depend on \pt{} and centrality range; minimum and maximum values are listed here. Empty fields indicate that a given check does not apply to the particle of interest. If an uncertainty has been tested but cannot be resolved within statistical precision, the field is marked \emph{negl} for negligible. Horizontal lines are used to separate the different categories of systematic uncertainties as explained in Sec.~\ref{sec:systematics}.} 
    \end{center}
\end{table}
\begin{table}[htbp]
    \begin{center}
        \begin{tabular*}{140mm}{@{\extracolsep{\fill}}lccc}
            Error source & \pipm{} & \kapm{} & p+\pbar{} \\
            \hline\hline
            Vertex position & 1--3\% & 1--3\% & 1--3\% \\
            1\% wide centrality intervals & 0--1\% & 0--1\% & 0--1\% \\
            Centrality estimator & 1--3\% & 1--3\% & 2--3\% \\
            Magnetic field polarity & 1--2\% & 1--3\% & 1--3\% \\
            Interaction rate & 1--2\% & 2--3\% & 2--3\% \\
            Pileup rejection & 0--2\% & 1--2\% & 2--3\% \\
            \hline
            Tracking mode & 0--2\% & 1--5\% &  1--10\% \\
            Number of TPC space points & 0--1\% & 0--1\% & 0--1\% \\
            Track quality & 3--4\% & 2--3\% & 3--4\% \\
            \hline
            Particle identification purity & 1--4\% & 2--4\% & 2--5\% \\
            Number of TPC clusters used for \dEdx{} & 0--2\% & 0--1\% & 0--1\% \\
            Exclusive particle identification & 0--1\% & 0--2\% & 0--1\% \\
            \hline 
            Positive and negative rapidities & 1--3\% & 1--2\% & 2--3\% \\
            Opposite charges & 2--3\% & 2--3\% & 2--3\% \\
            $v_{\rm n}$ from V0A or V0C & 1--3\% & 2--4\% & 2--4\% \\
            Channel removal from V0A & 6--14\% & 6--14\% & 5--15\% \\
            \hline
   \end{tabular*}
   \caption{\label{tab:systematics3} Summary of systematic uncertainties for the $v_4$ of \pipm{}, \kapm{}, and p+\pbar{}. The uncertainties depend on \pt{} and centrality range; minimum and maximum values are listed here. Horizontal lines are used to separate the different categories of systematic uncertainties as explained in Sec.~\ref{sec:systematics}.} 
    \end{center}
\end{table}

\subsubsection*{Event selection}

The nominal event selection criteria and centrality determination are discussed in Sec.~\ref{sec:evtrack}. Event selection criteria are varied 
by (i) changing the default centrality estimator from energy deposition in the V0 scintillator to either an estimate based on the number of hits 
in the first or second layer of the ITS; (ii) performing the $v_{\rm n}$ analysis of \pipm{}, \kapm{}, and p+\pbar{} in 1\% wide centrality intervals 
to test the effect of multiplicity fluctuations (a test not possible for \kanull{}, \lambdas{} $v_3$); (iii) not rejecting events with 
tracks caused by pileup or imposing a stricter than default pileup rejection by requiring a tighter correlation between the V0 and central 
barrel multiplicities; (iv) requiring the reconstructed primary vertex of a collision to lie alternatively within $\pm$12~cm and $\pm$5~cm 
from the nominal interaction point along the beam axis; (v) analyzing events recorded under different magnetic field polarities 
independently; (v) analyzing events recorded at different collision rates independently. 

\subsubsection*{Charged particle tracking}

The nominal charged particle track selection criteria are outlined in Sec.~\ref{sec:evtrack}. Charged particle track selection criteria are varied 
by (i) requiring the third layer of the ITS to be part of the track reconstruction rather than the first two layers only; (ii) using only tracks that have at least 
three hits per track in the ITS, complemented by tracks without hits in the first two layers of the ITS (in which case the primary interaction vertex 
is used as an additional constraint for the momentum determination); (iii) changing the requirement on the minimum number of TPC space points 
that are used in the reconstruction from 70 to 60, 80, and 90; (iv) an additional systematic uncertainty is evaluated combining the following checks of 
the track quality: rejecting tracks that are reconstructed close to the sector boundaries of the TPC to which the sensitive pad rows do not extend, varying the minimum number of crossed TPC pad rows from 70 to 120, and requesting at least 90\% instead of 80\% of the 
expected TPC space points to reconstruct a track. Variations (i) and (ii) are referred to as \emph{tracking mode} in Tabs.~\ref{tab:systematics1}, \ref{tab:systematics2}, and \ref{tab:systematics3}. 

\subsubsection*{Particle identification}

The nominal particle identification approach for \pipm{}, \kapm{}, and p+\pbar{} is outlined in Sec.~\ref{sec:pid}. Particle identification criteria are 
varied by (i) changing the minimum number of clusters in the TPC that are used to estimate the \dEdx{} from 70 to 60, 80, and 90; (ii) rejecting tracks that 
satisfy the particle identification criterion for more than one particle species simultaneously for \pt{} $<$ 4~\GeVc{}; (iii) varying the particle 
identification criterion from ${\rm n}\sigma_{\rm PID} < 3$ to ${\rm n}\sigma_{\rm PID} < 1$, ${\rm n}\sigma_{\rm PID} < 2$, and 
${\rm n}\sigma_{\rm PID} < 4$; (iv) varying the ${\rm n}\sigma_{\rm TPC}$ ranges that are used for particle identification for \pt{} $>$ 4~\GeVc{}. 

\subsubsection*{The \vo{} finding and $\phi$-meson reconstruction}

The nominal \vo{} finding strategy is described in Sec.~\ref{sec:v0s}. The \vo{} finding criteria fall into two categories: topological requirements 
on the \vo{}s themselves, and selection imposed on their daughter tracks. These criteria are varied by (i) requiring a minimum 
\pt{} of the \vo{} daughter tracks of 0.2 \GeVc{}; (ii) changing the requirement on the minimum number of TPC space points that are used in 
the reconstruction of the \vo{} daughter tracks form 70 to 60 and 80; (iii) varying the minimum number of TPC padrows crossed by the \vo{} 
daughter tracks from 70 to 60 and 80; (iv) requesting at least 90\% or 70\% instead of 80\% of the expected TPC space points to reconstruct 
the \vo{} daughter tracks; (v) changing the maximum DCA of the \vo{} daughter tracks to the secondary vertex from 0.5~cm to 0.3~cm and 0.7~cm; 
(vi) changing the minimum DCA of the \vo{} daughter tracks to the primary vertex from 0.1~cm to 0.05~cm and 0.3~cm; (vii) varying the number 
of clusters in the TPC that are used to estimate the \dEdx{} of the \vo{} daughter tracks from 70 to 60 and 90; (viii) varying the particle 
identification criterion of the \vo{} daughter tracks from $|{\rm n}\sigma_{\rm TPC}| < 3$ to $|{\rm n}\sigma_{\rm TPC}| < 1$ and 
$|{\rm n}\sigma_{\rm TPC}| < 4$; (ix) changing the minimum value of the $\cos \theta_p$ from 0.998 to 0.98; (x) varying the minimum 
radial distance to the primary vertex at which the \vo{} can be produced from 5~cm to 1~cm and 15~cm; (xi) varying the maximum radial distance 
to the beam pipe at which the \vo{} can be produced from 100~cm to 50~cm and 150~cm; (xii) the contamination from \lambdas{} decays 
and $\gamma$-conversions to the \kanull{} sample is checked by only selecting \vo{} daughter tracks with a \dEdx{} value 2$\sigma$ away 
from the expected electron \dEdx{}, effectively excluding electrons, and limiting the value of the Armenteros-Podolanski variables $\alpha$ and $q$.

The yield extraction, as explained in Sec~\ref{sec:v0s} for the \kanull{} and \lambdas{}, and Sec~\ref{sec:phi} for the $\phi$-meson, is varied by: 
(i) using a third-order polynomial as parametrization of residual background in the invariant mass spectra; (ii) using for the $\phi$-meson
a Voigtian distribution (a convolution of a Gaussian distribution and Breit-Wigner distribution, where the width of the Breit-Wigner distribution is 
set to the natural width of the $\phi$-meson, allowing for the Gaussian distribution to describe the smearing of the $\phi$-meson width due to the 
detector resolution) for the parametrization of the $\phi$-meson invariant mass yield; using for the \kanull{} and \lambdas{} a sum of two Gaussian 
distributions with the same mean for the parametrization of the \kanull{}, \lambdas{} invariant mass yield; (iii, for the $\phi$-meson only) using the yield 
of like-sign kaon pairs, in which two kaons with equal charge from the same event are used as candidate, for background yield description instead of 
event mixing. 

\subsubsection*{Extraction of the \vnpt{}}

The nominal approach of measuring \vnpt{} is outlined in Sec.~\ref{sec:flow}, and is varied by: (i) performing flow analysis for \pipm{}, \kapm{}, 
and p+\pbar{} for positive and negative charges independently; (ii) performing flow analysis for positive and negative rapidities 
independently; (iii) performing flow analysis for \pipm{}, \kapm{}, and p+\pbar{} in 1\% centrality 
intervals and merging the result rather than measuring in wider centrality intervals directly; (iv) suppressing the signal from a specific 
V0A channel in the evaluation of the \textbf{Q}$_{\rm n}$-vector (see Eq.~\ref{eq:Qcomp}), which, on average, measures a lower energy 
deposition with respect to the ones reported by the other channels from the same ring; (v) performing flow analysis with the 
${\bf Q}_{\rm n}$-vector calculated from the V0A or V0C separately; (vi) testing various M$_{\rm d^+ d^-}$ intervals over which 
$v_{n}^{\rm{bg}}({\rm{M}_{\rm{d}^+\rm{d}^-}})$ is fitted; (vii) testing the assumption 
made on $v_{\rm n}^{\rm bg}$ by changing the parametrization from a second-order polynomial to a first-order polynomial function.

\section{Results and discussion}
\label{sec:results}

The flow coefficients $v_2$, $v_3$, and $v_4$ of identified particles are presented for various centrality classes in Sec.~\ref{sec:rescendep}; scaling properties are discussed in Sec.~\ref{sec:resscaling}. Comparisons to various model calculations, studies on the shape 
evolution of \vnpt{} with centrality, and comparisons to $v_{\rm n}$ measured at \sqrtSnn{} = 2.76 TeV are shown in 
Secs.~\ref{sec:reshydro},~\ref{sec:resshape}, and~\ref{sec:resrun1}, respectively.

\subsection{Centrality and \pt{} dependence of flow coefficients}
\label{sec:rescendep}
\begin{figure}
    \includegraphics[width=\textwidth]{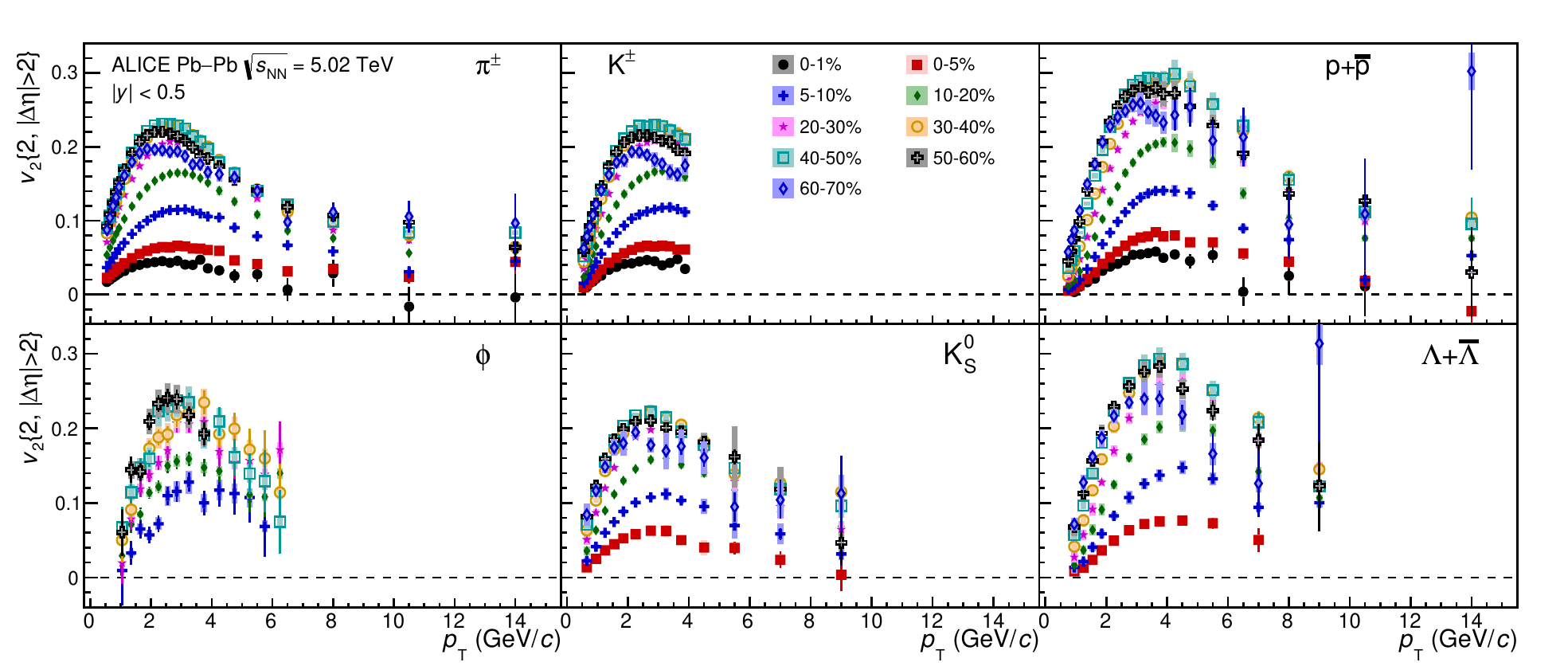}
    \caption{(Colour online) Centrality dependence of \vtwopt{} for \pipm{}, \kapm{}, p+\pbar{}, \lambdas{}, \kanull{}, and the $\phi$-meson. Statistical and systematic uncertainties are shown as bars and boxes, respectively.}
    \label{fig:v2pidcen}
\end{figure}
\begin{figure}
    \includegraphics[width=\textwidth]{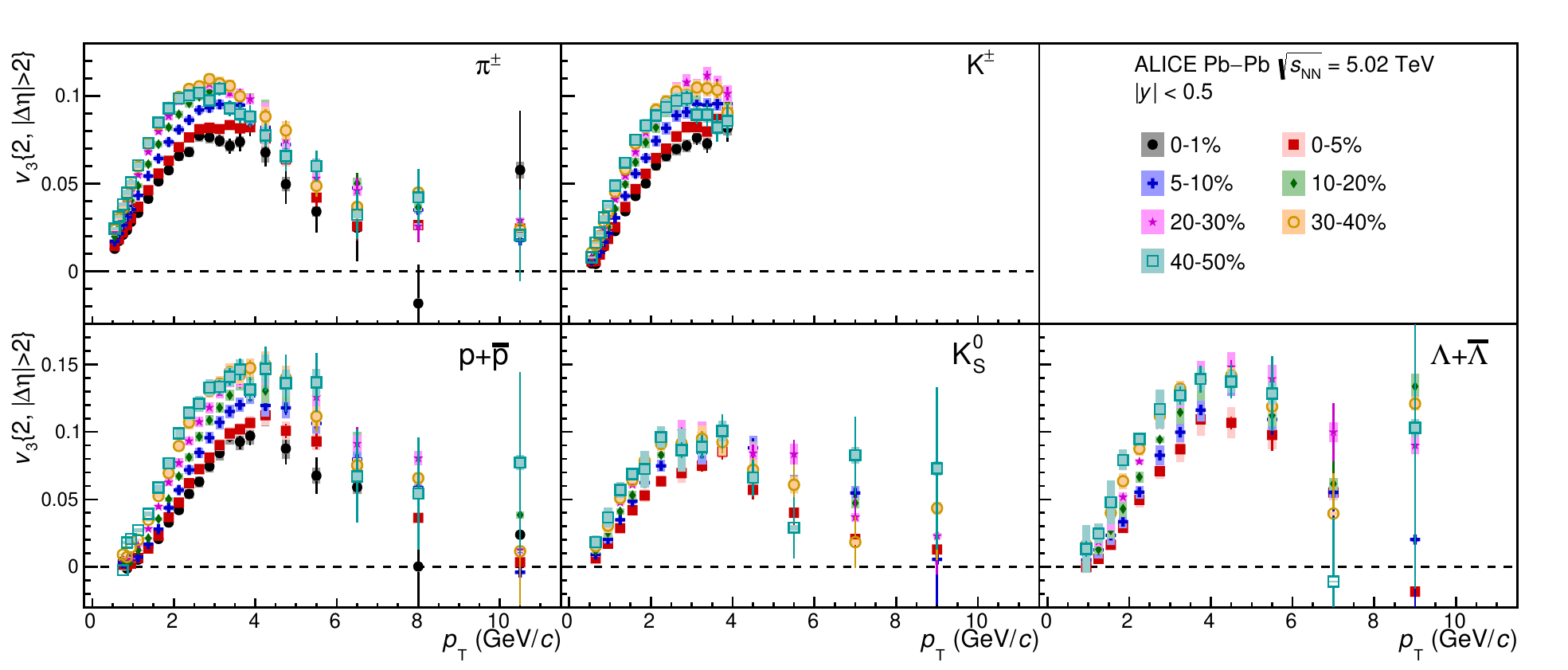}
    \caption{(Colour online) Centrality dependence of \vthreept{} for \pipm{}, \kapm{}, p+\pbar{}, \lambdas{}, and \kanull{}. Statistical and systematic uncertainties are shown as bars and boxes, respectively.}
    \label{fig:v3pidcen}
\end{figure}
\begin{figure}
    \includegraphics[width=\textwidth]{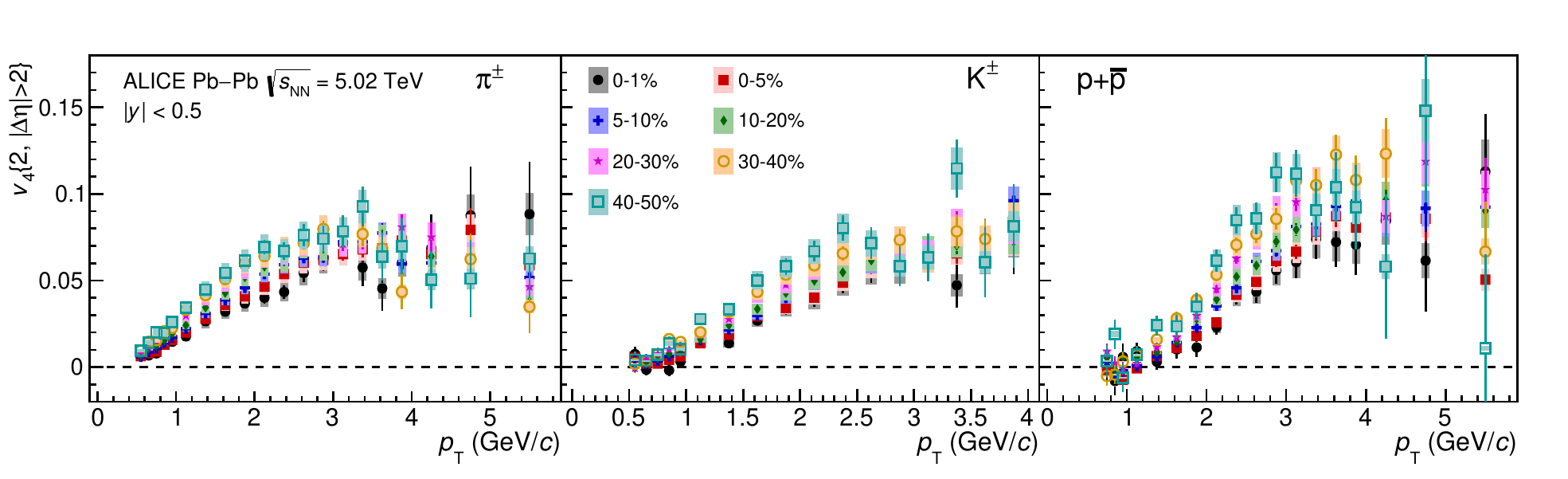}
    \caption{(Colour online) Centrality dependence of \vfourpt{} for \pipm{}, \kapm{}, and p+\pbar{}. Statistical and systematic uncertainties are shown as bars and boxes, respectively.}
    \label{fig:v4pidcen}
\end{figure} 

Figure~\ref{fig:v2pidcen} shows the \vtwopt{} of \pipm{}, \kapm{}, p+\pbar{}, \lambdas{}, \kanull{}, and the $\phi$-meson for various 
centrality classes in the range 0--70\%. For the \pipm{}, \kapm{} and p+\pbar{}, measurements performed in ultra-central collisions 
(0--1\%) are also presented. For the $\phi$-meson, the results are reported in the 5--60\% centrality range, where $v_2$ can be
measured accurately. The magnitude of $v_2$ increases strongly with decreasing centrality up to the 40--50\% centrality interval 
for all particle species. This 
evolution is expected, since the eccentricity of the overlap zone of the colliding nuclei increases for peripheral collisions and $v_2$ scales 
approximately linearly with eccentricity~\cite{Gardim:2011xv}. For more peripheral collisions (i.e.\ 50--60\% and 60--70\%), the value of $v_2$ 
is smaller than in the previous centrality intervals for all particle species except the $\phi$-meson. This suggests that the system has a shorter 
lifetime in more peripheral collisions, which does not allow for the generation of large $v_2$ \cite{Song:2007fn}. Furthermore, the reduced 
contribution of eccentricity fluctuations and hadronic interactions might play an important role in these centrality ranges \cite{song}. A 
non-zero, positive $v_2$ is found in the 0--1\% centrality interval for $\pt<6$ \GeVc{} for \pipm{}, \kapm{}, and p+\pbar{}, which mostly reflects the 
contribution from event-by-event fluctuations in the initial nucleon and gluon density as the system shape is almost spherical at vanishing impact parameter. 

The third-order flow coefficent $v_3$ is generated by inhomogeneities in the initial nucleon and gluon density and not by the collision geometry \cite{fluc1,fluc2,fluc3,fluc4}, while $v_4$ arises from initial collision geometry, fluctuations, and the non-linear hydrodynamic response of the medium \cite{nl1,nl2}. Higher-order flow harmonics are more sensitive to transport coefficients than $v_2$ \cite{fluc4}, as the dampening effect of $\eta/s$ 
leads to a stronger decrease of these coefficients \cite{sens1,sens2}. Figures~\ref{fig:v3pidcen} and \ref{fig:v4pidcen} present the 
\vthreept{} of \pipm{}, \kapm{}, p+\pbar{}, \lambdas{}, and \kanull{} and \vfourpt{} of \pipm{}, \kapm{}, and p+\pbar{} for various centrality classes in the 0--50\% range. Statistical precision limits extending the $v_4$ measurement to more peripheral collisions or carrying it out for \lambdas{}, \kanull{}, and the $\phi$-meson. Non-zero, positive $v_3$ and $v_4$ are observed for particle species throughout the entire \pt{} ranges up 
to $\approx$ 8 \GeVc{}. Unlike $v_2$, the coefficients $v_3$ and $v_4$ increase weakly from ultra-central to peripheral collisions. This observation 
illustrates that higher-order flow coefficients are mainly generated by event-by-event fluctuations in the initial nucleon and gluon density. 

All flow coefficients increase monotonically with increasing \pt{} up to 3-4~\GeVc{} where a maximum is reached. The position of this 
maximum depends on centrality and particle species as it takes place at higher \pt{} for heavier particles for various centrality classes. This 
behaviour can be explained by the centrality dependence of radial flow combined with the parton density, which will be detailed in Sec.~\ref{sec:resshape}.

\begin{sidewaysfigure}  
  \centering
    \includegraphics[width=0.9\textwidth,  height=13cm]{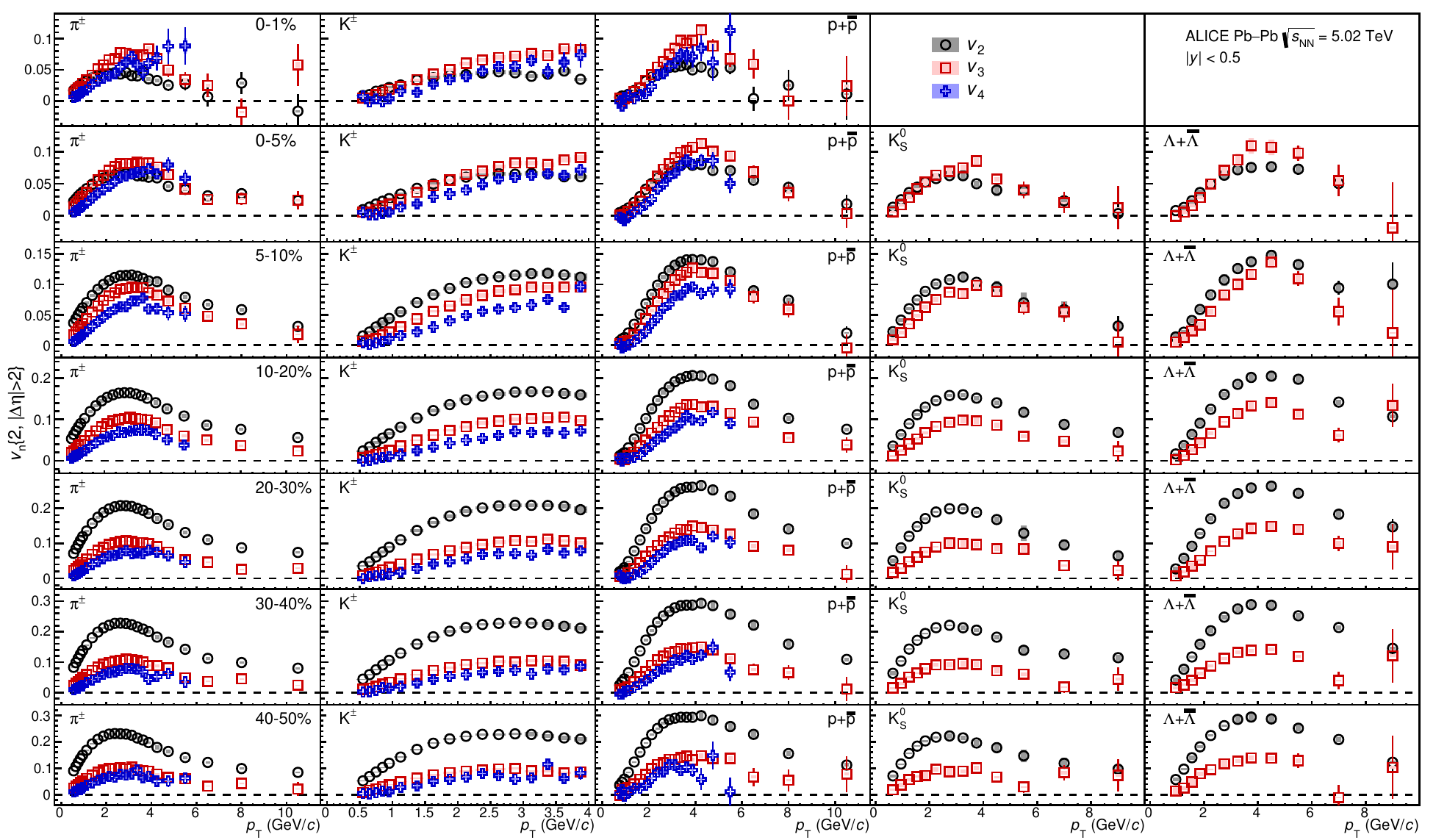}
    \caption{(Colour online) The evolution of \vnpt{} of \pipm{}, \kapm{}, p+\pbar{}, \lambdas{}, and \kanull{} for various centrality classes. Statistical and systematic uncertainties are shown as bars and boxes, respectively.}
     \label{fig:vn_piKpK0L_V0A}
\end{sidewaysfigure}

Figure~\ref{fig:vn_piKpK0L_V0A} presents the evolution of \vnpt{} of different particle species for various centrality classes. In the most central 
collisions, initial nucleon-density fluctuations are expected to be the main contributor to the generation of $v_{\rm n}$. For the 0--1\% centrality 
interval, $v_3$ is the dominant flow coefficient for $1.5<\pt<6.0$ \GeVc{}, $2.0<\pt<4$ \GeVc{}, and $2.5<\pt<6$ \GeVc{} for \pipm{}, 
\kapm{}, and p+\pbar{}, respectively. Furthermore, $v_4$ becomes equal to $v_2$ at $\pt \approx$ 2.0 \GeVc{} (2.2, 2.5) for \pipm{} (\kapm{}, p+\pbar{}), after which it increases gradually and reaches a magnitude similar to $v_3$ at around 
3.5~GeV/$c$. A similar trend is observed in the 0--5\% centrality class for all particle species. However, the crossing between flow coefficients 
(the \pt{} value at which they reach a similar magnitude), which also depends on the particle mass, takes place at different \pt{} values 
than for the 0--1\% centrality interval. This dependence of the crossing between different flow coefficients can be attributed to the interplay of elliptic, triangular, and quadrangular flow with radial flow. Upwards of 5\% collision centrality, $v_2$ is larger than $v_3$ and $v_4$, confirming 
the hypothesis that collision geometry dominates the generation of flow coefficients.  

\begin{figure}
    \includegraphics[width=\textwidth]{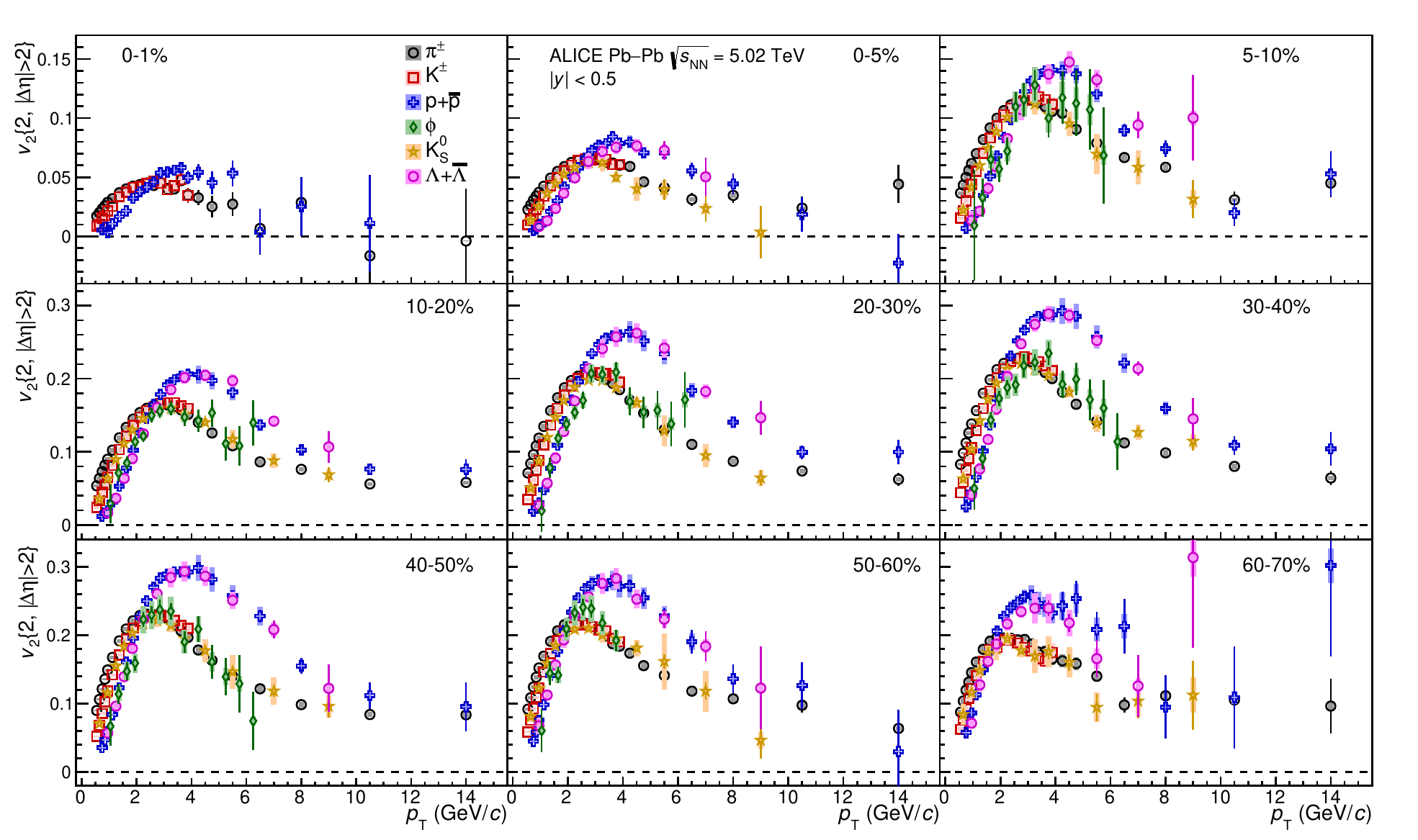}
    \caption{(Colour online) The \pt-differential $v_2$ of \pipm{}, \kapm{}, p+\pbar{}, \lambdas{}, \kanull{}, and the $\phi$-meson for various centrality classes. Statistical and systematic uncertainties are shown as bars and boxes, respectively.}
    \label{fig:v2pid}
\end{figure}

Figure~\ref{fig:v2pid} shows the \vtwopt{} of \pipm{}, \kapm{}, p+\pbar{}, \lambdas{}, \kanull{}, and the $\phi$-meson in a given centrality 
interval arranged into panels of various centrality classes, which allows for further illustration of the interplay between elliptic and 
radial flow. For $\pt<$ 2-3~\GeVc{}, $v_2$ of the different particle species is mass-ordered, meaning that lighter particles have a larger $v_2$ 
than heavier particles at the same \pt{}. This behaviour is indicative of strong radial flow which imposes an equal, isotropic velocity boost to all particles in 
addition to the anisotropic expansion of the medium \cite{massdep1, massdep2, massdep3}. For $3<\pt<$ 8-10~\GeVc{}, particles are grouped 
according to their number of constituent quarks, which supports the hypothesis of particle production via quark coalescence~\cite{dud2}. Particle 
type scaling and mass ordering are most directly tested by the $\phi$-meson $v_2$, as its mass is close to the proton mass. 
Figure~\ref{fig:v2pid} demonstrates that the $\phi$-meson $v_2$ follows proton $v_2$ at low \pt{}, but pion $v_2$ at intermediate \pt{} in all 
centrality classes. The crossing between meson and baryon $v_2$, which depends on the particle species, happens at higher \pt{} values 
for central than peripheral collisions as a result of the larger radial flow in the former. Lastly, it is seen that the $v_2$ of baryons is higher than 
that of mesons up to \pt{} $\approx$ 10 \GeVc{}, indicating that particle type dependence of $v_2$ persists up to high \pt{}. For $\pt>10$~\GeVc{}, 
where $v_2$ depends only weakly on transverse momentum, the magnitude of p+\pbar{} $v_2$ is compatible with that for \pipm{} within 
statistical and systematic uncertainties. Furthermore, the nuclear modification factor in this high \pt{} region is found to be the same for the 
two particle species within uncertainties~\cite{Adam:2015kca}.  

\begin{figure}
    \includegraphics[width=\textwidth]{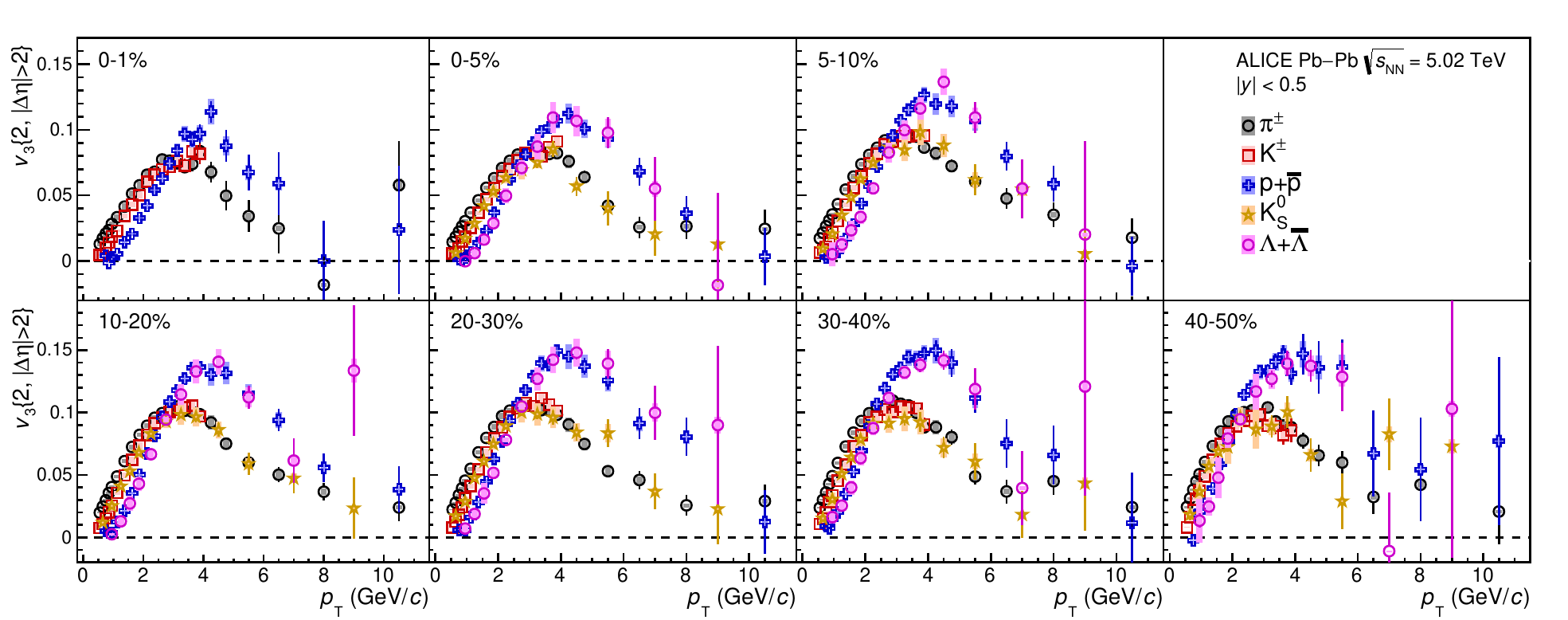}
    \caption{(Colour online) The \pt-differential $v_3$ of \pipm{}, \kapm{}, p+\pbar{}, \lambdas{}, and \kanull{} for various centrality classes. Statistical and systematic uncertainties are shown as bars and boxes, respectively.}
    \label{fig:v3pid}
\end{figure}

\begin{figure}
    \includegraphics[width=\textwidth]{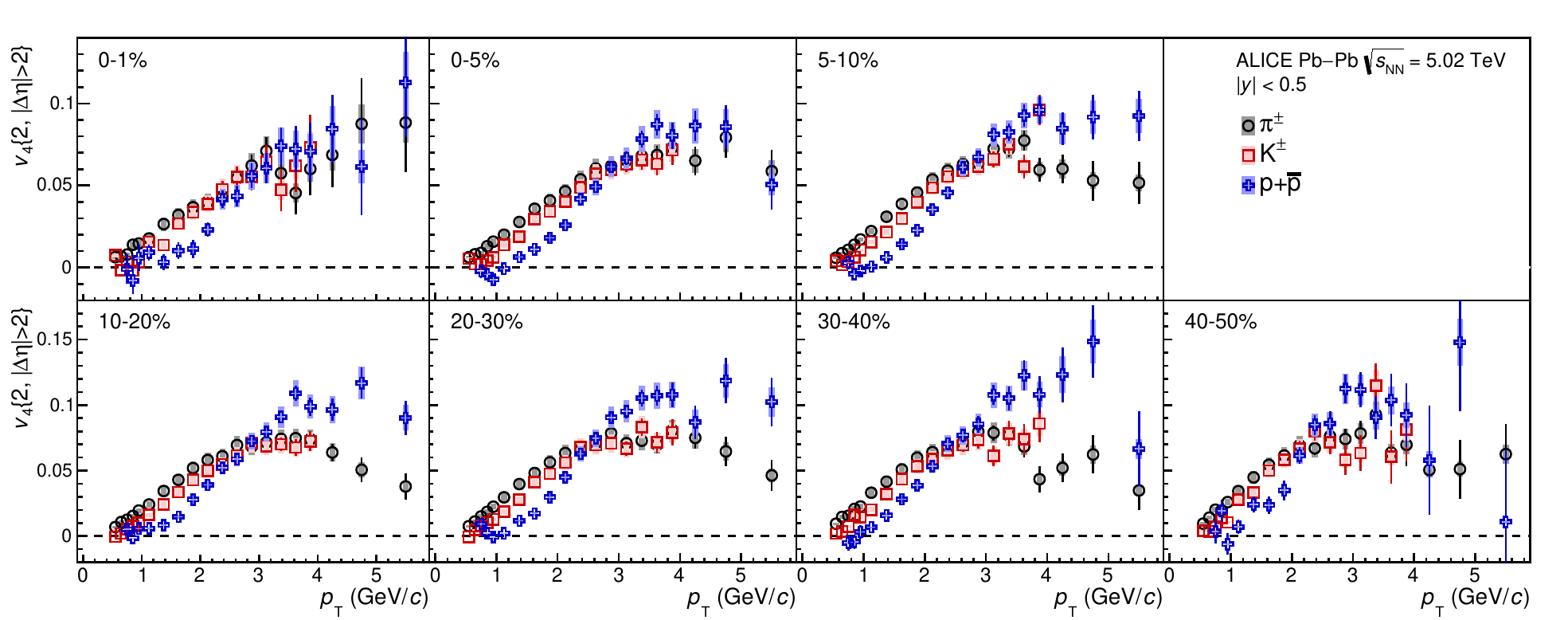}
    \caption{(Colour online) The \pt-differential $v_4$ of \pipm{}, \kapm{}, and p+\pbar{} for various centrality classes. Statistical and systematic uncertainties are shown as bars and boxes, respectively.} 
    \label{fig:v4pid}
\end{figure}

Figures~\ref{fig:v3pid} and~\ref{fig:v4pid} present the \vthreept{} and \vfourpt{} for different particle species in a given centrality 
interval. Both $v_3$ and $v_4$ show a clear mass ordering at $\pt<$ 2-3~\GeVc{}, confirming the interplay between 
triangular and quadrangular flow and radial flow. For $3<\pt<8$~\GeVc{}, particles are grouped into mesons and baryons and, analogous to the 
trend of $v_2$ in this \pt{} region, the flow of baryons is larger than that of mesons. The crossing between meson and baryon $v_3$ and 
$v_4$ also exhibits a centrality and particle mass dependence. 

Figures~\ref{fig:v2pid} and~\ref{fig:v3pid} also show a comparison between \kapm{} and \kanull{} $v_2$ and $v_3$ as a function 
of \pt{} for various centrality classes. A difference in \vnpt{} is found between the \kapm{} and \kanull{} measurements: the magnitude of \kanull{} 
$v_{\rm n}$ is systematically smaller than the magnitude of \kapm{} $v_{\rm n}$. This difference in $v_{\rm n}$ exhibits no \pt{} dependence, but 
changes with centrality for $v_2$. For $0.8 < \pt{} < 4.0$ \GeVc{}, the difference in $v_2$ ranges from 
$7\% \pm 3.5\%({\rm syst}) \pm 0.7\%({\rm stat})$ in the most central collisions to $1.5\% \pm 1.5\%({\rm syst}) \pm 0.4\%({\rm stat})$ in 
peripheral collisions. In the same kinematic range, a deviation in $v_3$ of $6.5\% \pm 5\%({\rm syst}) \pm 1.7\%({\rm stat})$ is found in the 
most central collisions and of $6\% \pm 4.5\%({\rm syst}) \pm 1\%({\rm stat})$ in peripheral collisions. This difference is similar in magnitude 
and centrality dependence as the one reported by ALICE in Pb--Pb collisions at \sqrtSnn{} = 2.76 TeV in  \cite{Abelev:2014pua}.

\subsection{Scaling properties}
\label{sec:resscaling}

\begin{figure}
    \includegraphics[width=\textwidth]{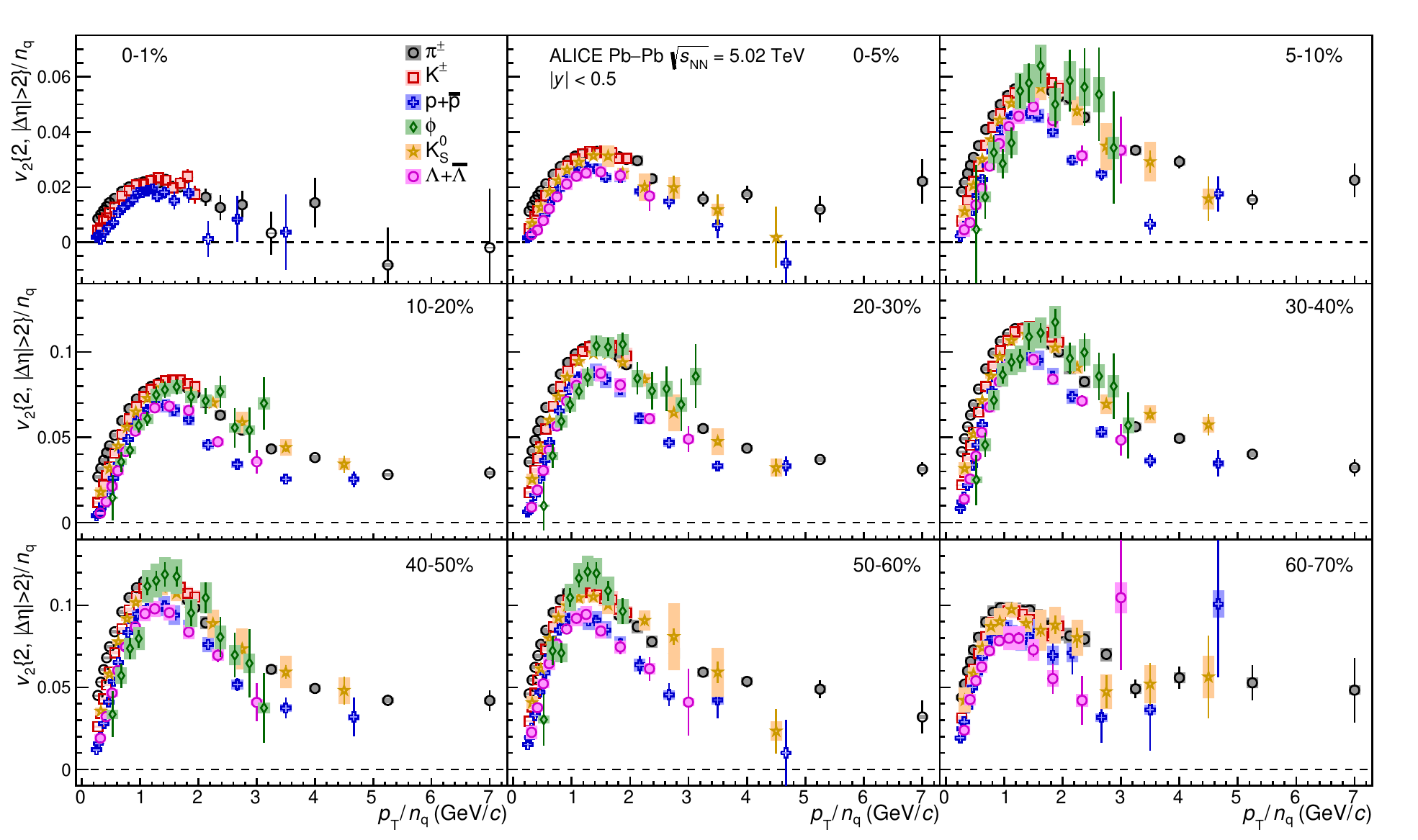}
    \caption{(Colour online) The $\pt/n_{\rm q}$ dependence of $v_2/n_{\rm q}$ of \pipm{}, \kapm{}, p+\pbar{}, \lambdas{}, \kanull{}, and the $\phi$-meson for various centrality classes. Statistical and systematic uncertainties are shown as bars and boxes, respectively.}
    \label{fig:v2pid_scaled}
\end{figure}

\begin{figure}
    \includegraphics[width=\textwidth]{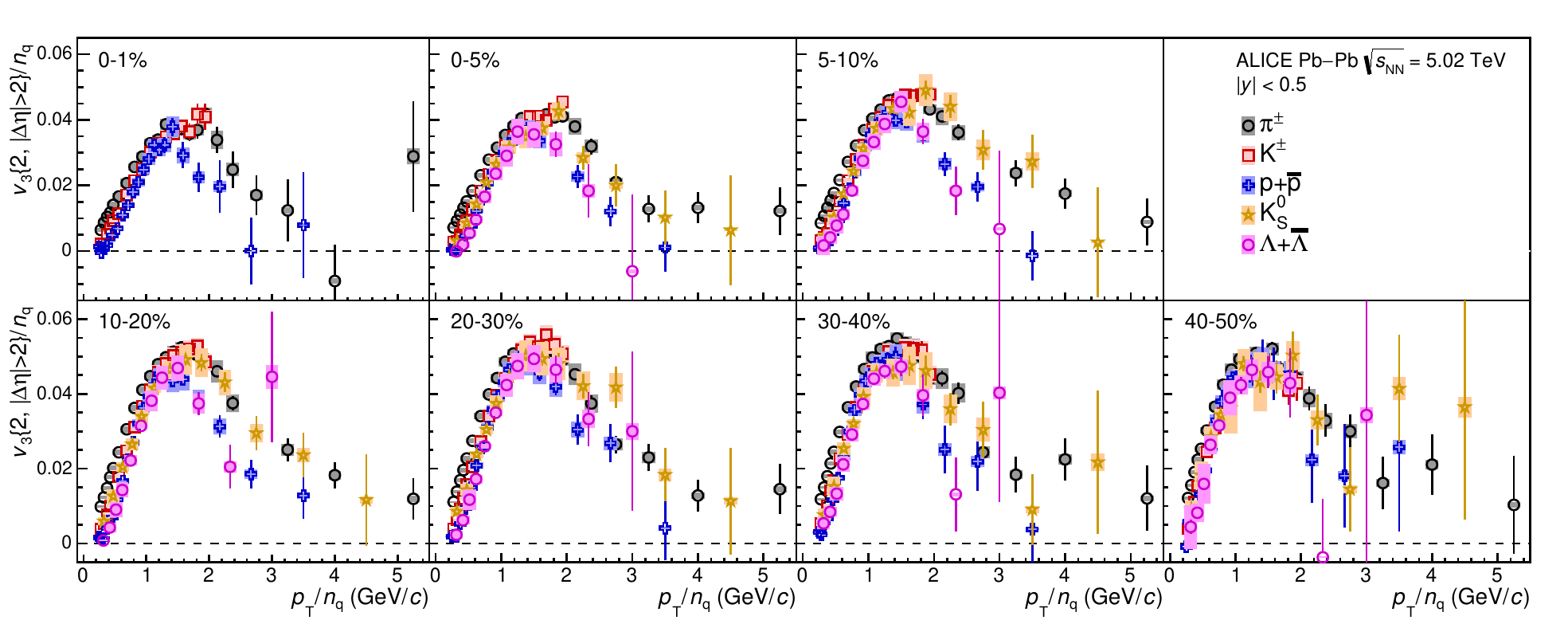}
    \caption{(Colour online) The $\pt/n_{\rm q}$ dependence of $v_3/n_{\rm q}$ of \pipm{}, \kapm{}, p+\pbar{}, \lambdas{}, and \kanull{} for various centrality classes. Statistical and systematic uncertainties are shown as bars and boxes, respectively.} 
    \label{fig:v3pid_scaled}
\end{figure}

\begin{figure}
    \includegraphics[width=\textwidth]{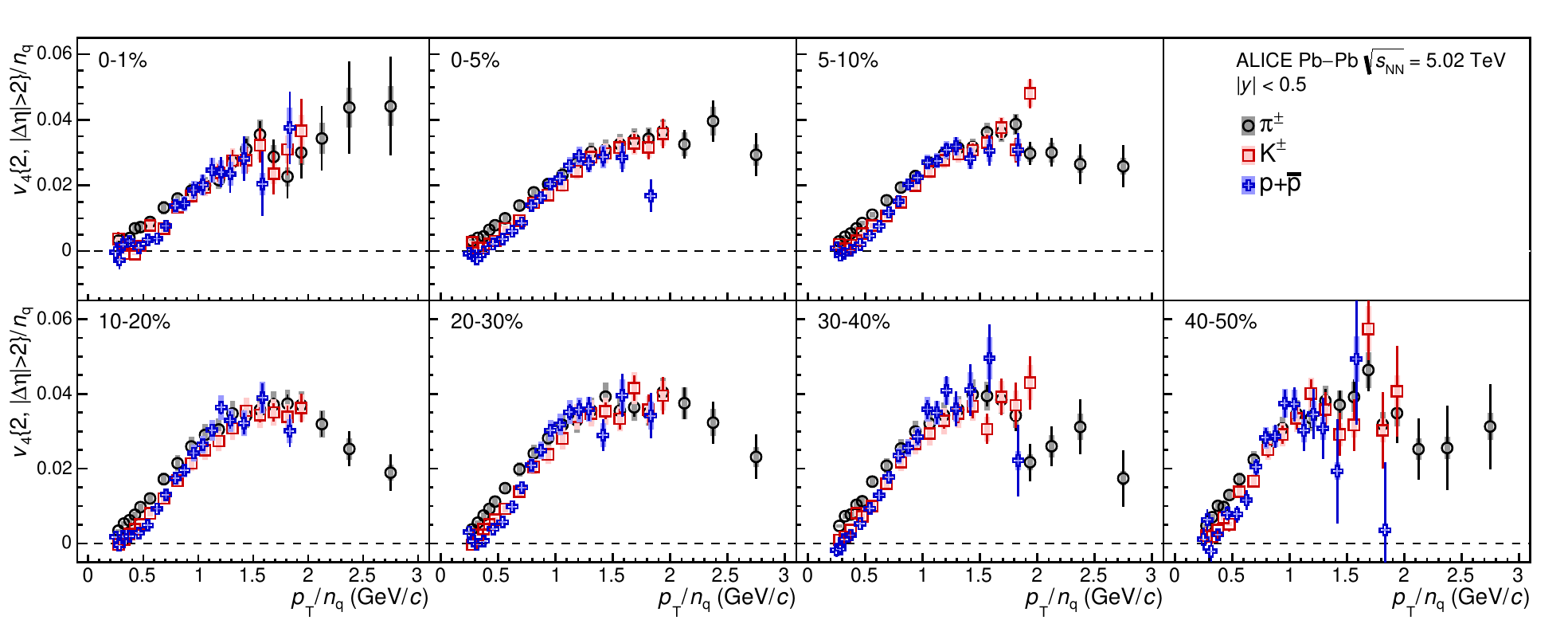}
    \caption{(Colour online) The $\pt/n_{\rm q}$ dependence of $v_4/n_{\rm q}$ of \pipm{}, \kapm{}, and p+\pbar{} for various centrality classes. Statistical and systematic uncertainties are shown as bars and boxes, respectively.} 
    \label{fig:v4pid_scaled}
\end{figure}

To test the hypothesis of particle production via quark coalescence \cite{dud2}, which would lead to a grouping of $v_{\rm n}$ of 
mesons and baryons at intermediate \pt{}, both 
$v_{\rm n}$ and \pt{} are divided by the number of constituent quarks ($n_{\rm q}$) independently for each particle species. The 
$v_{2}/n_{\rm q}$, $v_{3}/n_{\rm q}$, and $v_{4}/n_{\rm q}$ of \pipm{}, \kapm{}, p+\pbar{}, \lambdas{}, \kanull{}, and the $\phi$-meson, 
plotted as a function of $\pt/n_{\rm q}$, are reported in Figs.~\ref{fig:v2pid_scaled}, \ref{fig:v3pid_scaled}, and \ref{fig:v4pid_scaled} 
for various centrality classes. 

For \pt{}$/n_{\rm q} >$ 1 \GeVc{}, the scaling is only approximate. To quantify the degree to which the measurements deviate from the 
$n_{\rm q}$ scaling, the $\pt/n_{\rm q}$ dependence of 
$v_{\rm n}/n_{\rm q}$ has been divided by a cubic spline fit to the p+\pbar{} $v_{\rm n}/n_{\rm q}$. In the region where quark 
coalescence is hypothesized to be the dominant process ($\approx 1 < \pt/n_{\rm q} < 3$~\GeVc{})~\cite{dud2,massmot2}, a 
deviation from the exact scaling of $\pm$ 20\% is found for $v_2$ for central collisions, which decreases to $\pm 15\%$ for the most 
peripheral collisions. For higher harmonics, a $\pm 20\%$ deviation is found for all centrality classes. This deviation is in agreement 
with earlier observations~\cite{Adare:2012vq,Abelev:2014pua,highham}.

\subsection{Comparison with model calculations}
\label{sec:reshydro}
\begin{figure}
    \begin{center}
    \includegraphics[width=\textwidth]{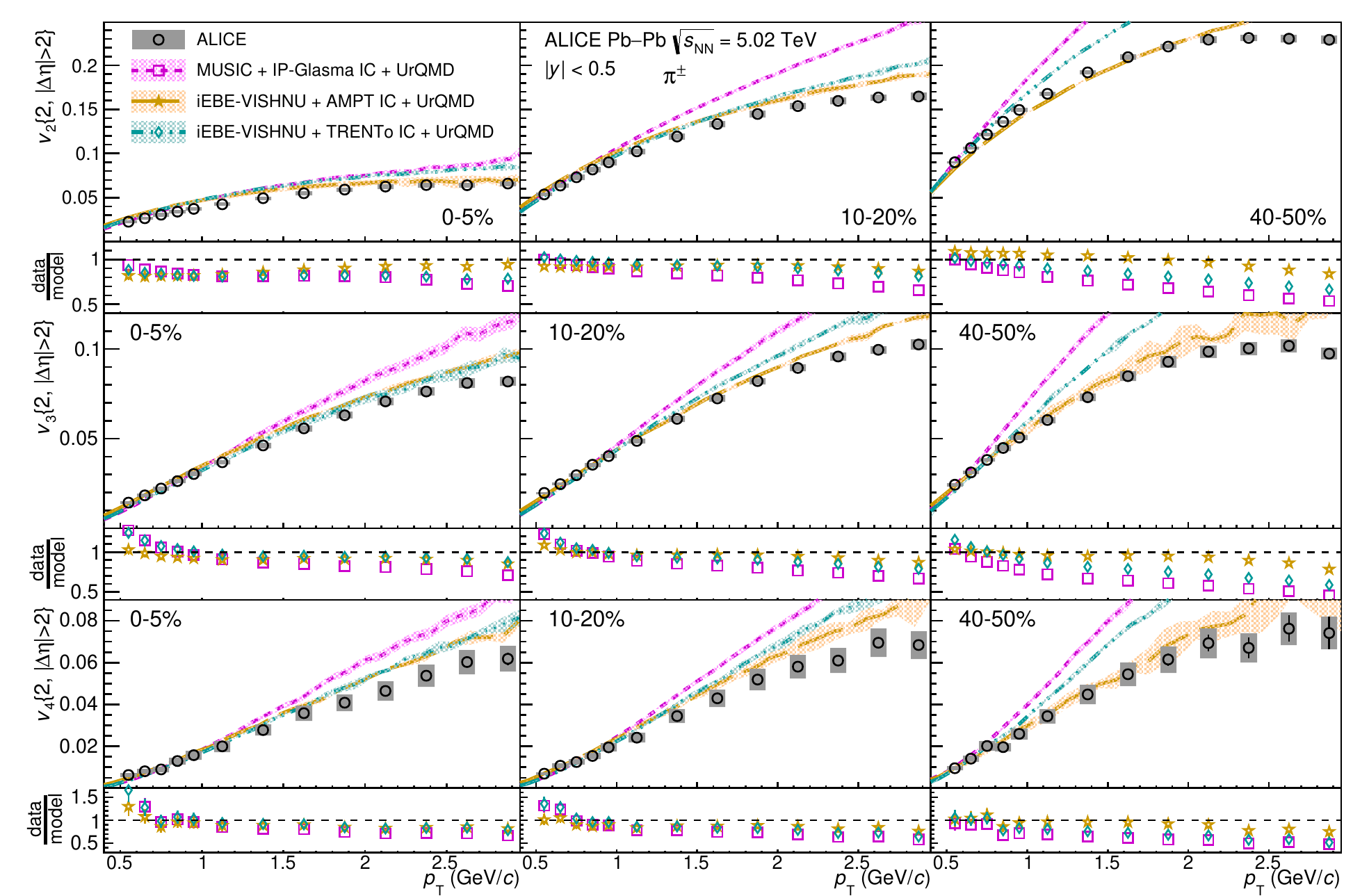}
    \caption{(Colour online) The $\pt$-differential $v_2$ (top), $v_3$ (middle), and $v_4$ (bottom) of \pipm{} for the 0--5\%, 10--20\%, and 40--50\% centrality classes compared to hydrodynamical calculations from MUSIC model using IP-Glasma initial conditions (magenta)~\cite{dud4} and the iEBE-VISHNU hybrid model using AMPT (orange) or \protect\trento{}~(cyan) initial conditions~\cite{wenbin}. Statistical and systematic uncertainties of the data points are shown as bars and boxes, respectively. The uncertainties of the hydrodynamical calculations are depicted by the thickness of the curves. The ratios of the measured $v_{\rm n}$ to a fit to the hydrodynamical calculations are also presented for clarity.}
    \label{fig:hydro1}
    \end{center}
\end{figure}
\begin{figure}
    \begin{center}
    \includegraphics[width=\textwidth]{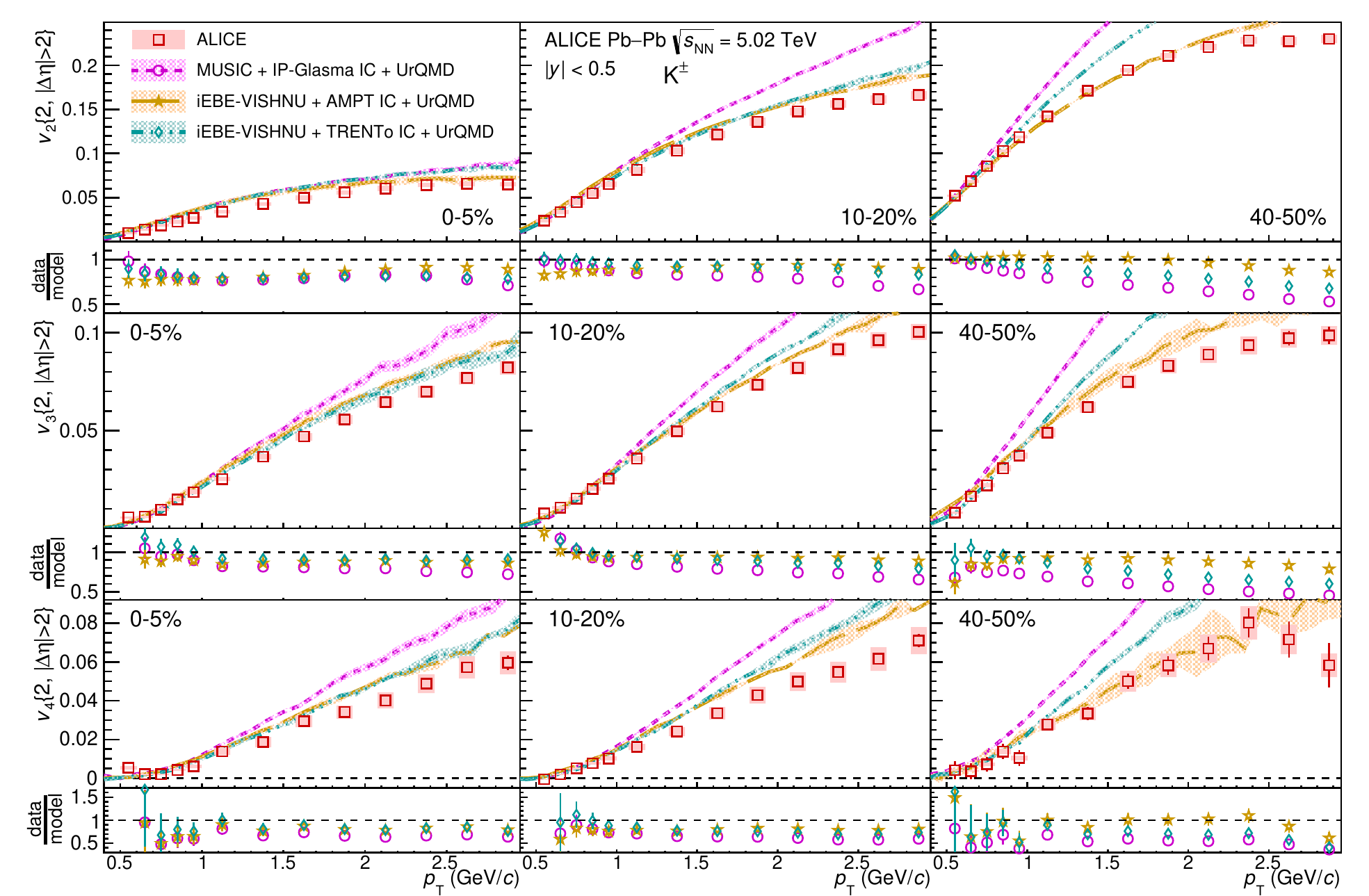}
    \caption{(Colour online) The $\pt$-differential $v_2$ (top), $v_3$ (middle), and $v_4$ (bottom) of \kapm{} for the 0--5\%, 10--20\%, and 40--50\% centrality classes compared to hydrodynamical calculations from MUSIC model using IP-Glasma initial conditions (magenta)~\cite{dud4} and the iEBE-VISHNU hybrid model using AMPT (orange) or \protect\trento{}~(cyan) initial conditions~\cite{wenbin}. Statistical and systematic uncertainties of the data points are shown as bars and boxes, respectively. The uncertainties of the hydrodynamical calculations are depicted by the thickness of the curves. The ratios of the measured $v_{\rm n}$ to a fit to the hydrodynamical calculations are also presented for clarity.}
    \label{fig:hydro2}
    \end{center}
\end{figure}
\begin{figure}
    \begin{center}
    \includegraphics[width=\textwidth]{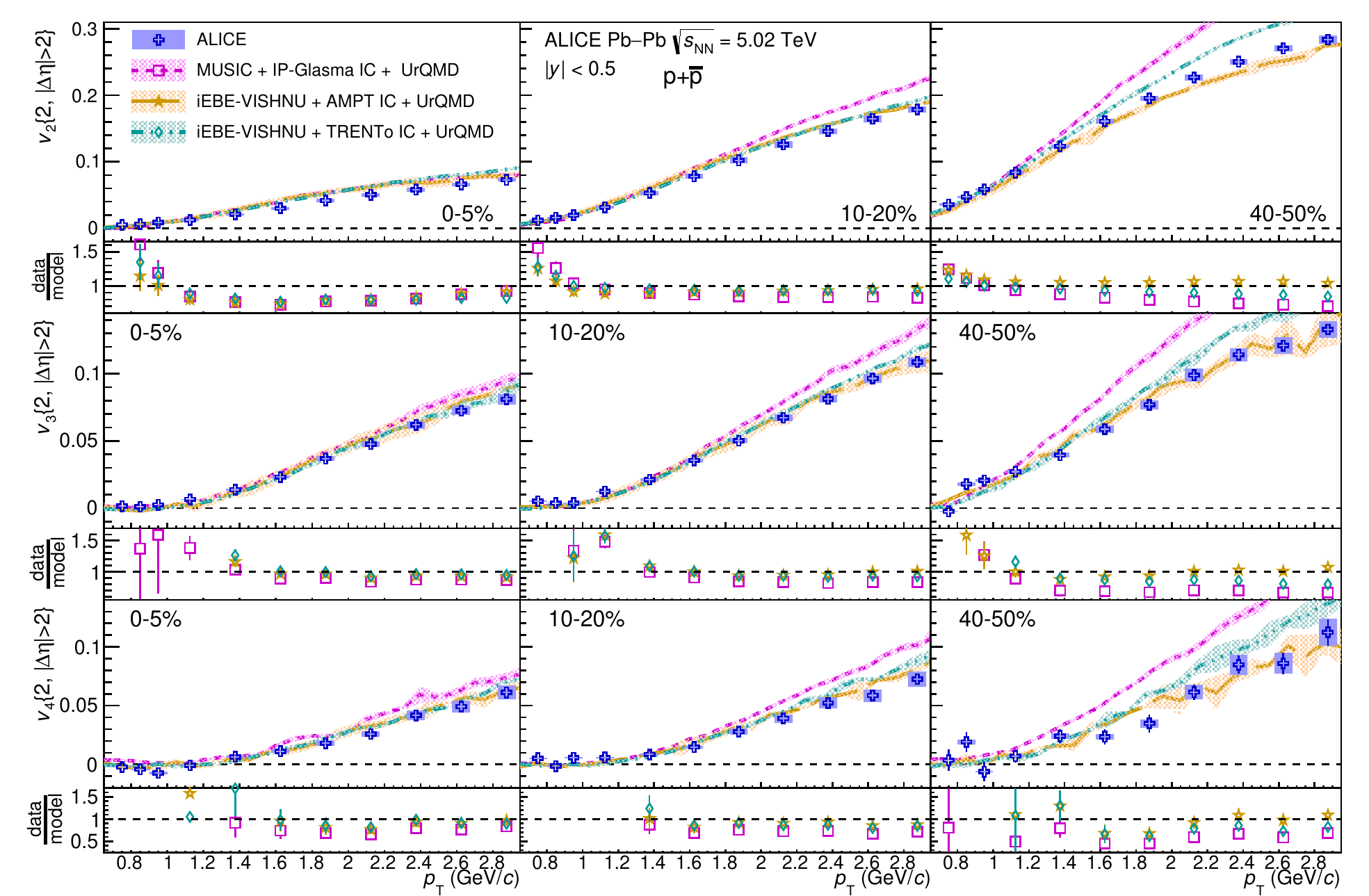}
    \caption{(Colour online) The $\pt$-differential $v_2$ (top), $v_3$ (middle), and $v_4$ (bottom) of p+\pbar{} for the 0--5\%, 10--20\%, and 40--50\% centrality classes compared to hydrodynamical calculations from MUSIC model using IP-Glasma initial conditions (magenta)~\cite{dud4} and the iEBE-VISHNU hybrid model using AMPT (orange) or \protect\trento{}~(cyan) initial conditions~\cite{wenbin}. Statistical and systematic uncertainties of the data points are shown as bars and boxes, respectively. The uncertainties of the hydrodynamical calculations are depicted by the thickness of the curves. The ratios of the measured $v_{\rm n}$ to a fit to the hydrodynamical calculations are also presented for clarity.}
    \label{fig:hydro3}
    \end{center}
\end{figure}
To test the validity of the hydrodynamic description of the QGP evolution, the $v_{\rm n}$ measurements in the 0--5\%, 
10--20\% and 40--50\% centrality intervals are compared to hydrodynamical calculations 
in Figs.~\ref{fig:hydro1},~\ref{fig:hydro2}, and~\ref{fig:hydro3} for \pipm{}, \kapm{}, and p+\pbar{}, respectively. Predictions from 
MUSIC~\cite{dud4} and iEBE-VISHNU~\cite{wenbin} simulations are depicted by the different 
coloured curves. The first calculation is based on MUSIC~\cite{Schenke:2010rr}, an event-by-event 3+1 dimensional viscous hydrodynamic 
model, coupled to a hadronic cascade model (UrQMD)~\cite{glasmacascade,bleichercascade}, which allows the influence of the 
hadronic phase on the anisotropic flow to be studied for different particle species. The IP-Glasma model~\cite{glasma1,glasma2} is used to simulate the 
initial conditions of the collision. MUSIC uses a starting time for the hydrodynamic evolution of $\tau_0=0.4$~fm/$c$, a switching 
temperature between the macroscopic hydrodynamic description and the microscopic transport evolution of $T_{\rm sw}=145$~MeV, a value 
of $\eta/s$ = 0.095, and a temperature dependent $\zeta/s$. The second calculation employs the iEBE-VISHNU hybrid 
model~\cite{Shen:2014vra}, which is an event-by-event version of the VISHNU hybrid model~\cite{Song:2010aq}, and couples 2+1 
dimensional viscous hydrodynamics VISH2+1~\cite{Song:2007fn} to UrQMD. The \trento{}~\cite{trento} and AMPT~\cite{ampt} models are used 
to describe the initial conditions. For both configurations, $\tau_0=0.6$~fm/$c$ and $T_{\rm sw}=148$~MeV are set from~\cite{Bernhard:2016tnd}, 
where these values have been obtained utilizing Bayesian statistics from a simultaneous fit of final charged-particle density, mean transverse 
momentum, and integrated flow coefficients $v_{\rm n}$ in Pb--Pb collisions at \sqrtSnn~=~2.76~TeV. The temperature-dependent 
$\eta/s$ and $\zeta/s$ extracted in~\cite{Bernhard:2016tnd} are used for \trento{}~initial conditions, while $\eta/s=0.08$ and 
$\zeta/s=0$ are taken for AMPT. 

Figures~\ref{fig:hydro1},~\ref{fig:hydro2}, and~\ref{fig:hydro3} show that the hydrodynamical calculations qualitatively reproduce the $v_{\rm n}$ 
measurements. The differences between the data points and models are visualized in Figs.~\ref{fig:hydro1},~\ref{fig:hydro2}, and~\ref{fig:hydro3} as the 
ratios of the measured $v_{\rm n}$ to a fit to the theoretical calculations. The iEBE-VISHNU calculations using AMPT initial conditions describe the 
\pt-differential $v_{\rm n}$ of \pipm{}, \kapm{}, and p+\pbar{} more accurately than \trento{}~based and MUSIC calculations for $\pt>1$~GeV/$c$. 
Using AMPT initial conditions, there is good agreement between \pipm{} and \kapm{} $v_{\rm n}$ and iEBE-VISHNU calculations for 
$\pt<2$~\GeVc{}, while p+\pbar{} $v_{\rm n}$ is described fairly well up to $\pt=3$~\GeVc{}. The \trento{}~based predictions follow 
\pipm{} and \kapm{} $v_{\rm n}$ up to slightly lower transverse momenta ($\pt<$1-2~\GeVc{}) and to $\pt{}<3$~\GeVc{} 
for p+\pbar{}, depending on the considered centrality interval. The MUSIC calculations are in agreement with the measured $v_{\rm n}$ 
for $\pt{} <1$~\GeVc{} in central collisions, however they overestimate $v_2$ at lower \pt{} in more peripheral collisions.

\subsection{Shape evolution of \vnpt{} as function of centrality}
\label{sec:resshape}

The evolution of the shape of \vnpt{} as function of centrality is quantified by taking the ratio of \vnpt{} in a given centrality interval to the \vnpt{} measured 
in the 20--30\% centrality interval
\begin{equation}
\label{eq:shape}
    \vnpt{}_{\rm ratio~to~20-30\%} = \frac{\vnpt{}}{\vnpt{}\vert_{20-30\%}}\frac{\vn{}\vert_{20-30\%}}{\vn{}},
\end{equation}
where the second fraction on the right-hand side of the equation serves as a normalization factor which is constructed from 
the \pt{}-integrated \vn{} values obtained in the 20--30\% centrality interval (\vn{}$\vert_{20-30\%}$) and the centrality interval 
of interest (\vn{}). Centrality-dependent variations in the shape of 
\vnpt{} will present themselves as deviations from unity of the observed $\vnpt{}_{\rm ratio~to~20-30\%}$.

\begin{figure}[!t]
    \begin{center}
    \includegraphics[width=\textwidth]{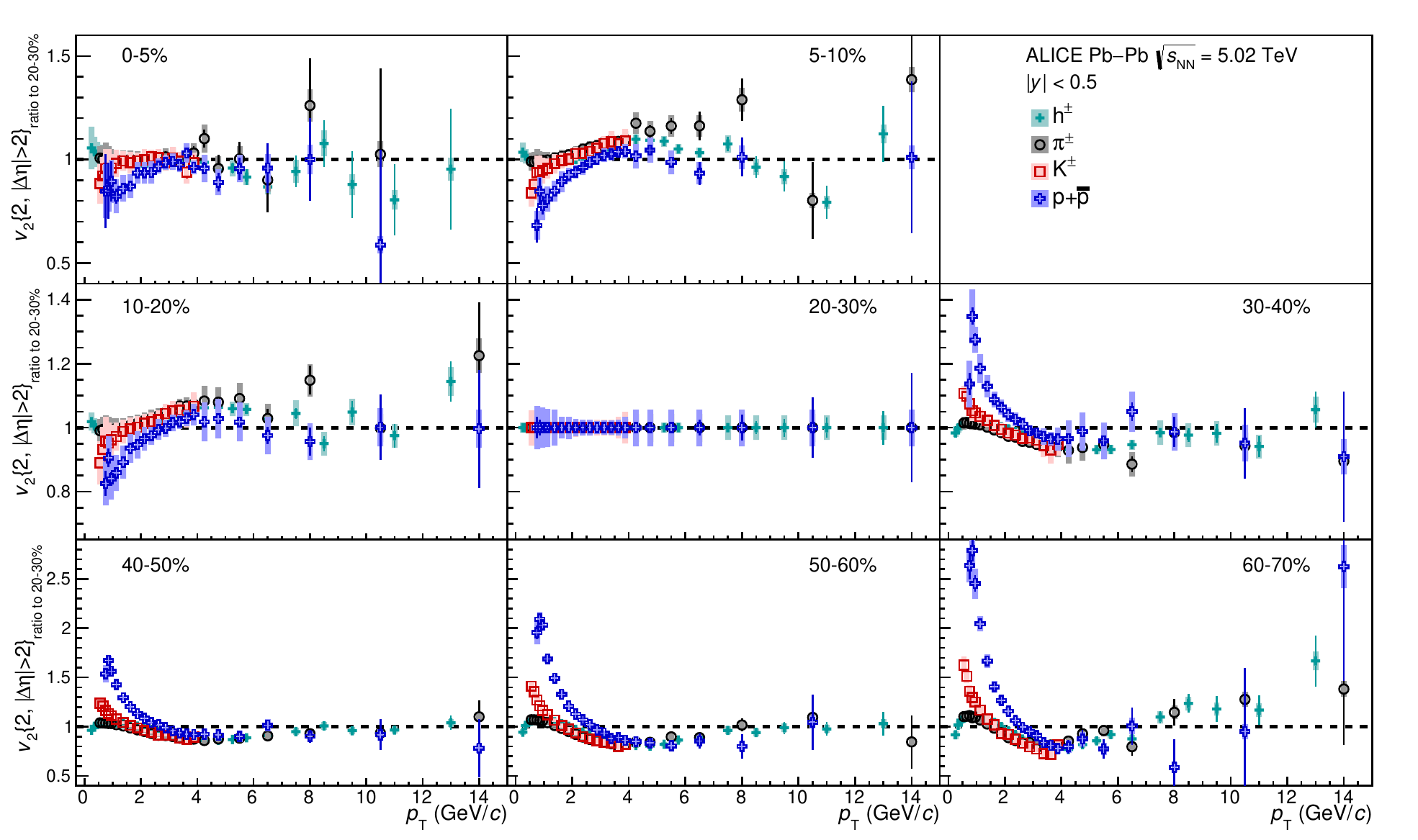}
    \caption{(Colour online) Centrality dependence of $\vtwopt{}_{\rm ratio~to~20-30\%}$ for \pipm{}, \kapm{}, p+\pbar{}, and inclusive charged particles \cite{Acharya:2018lmh}. Statistical and systematic uncertainties are shown as bars and boxes, respectively.}
    \label{fig:HistV2PtPIDRat2030.pdf}
    \end{center}
\end{figure}

\begin{figure}[!t]
    \begin{center}
    \includegraphics[width=\textwidth]{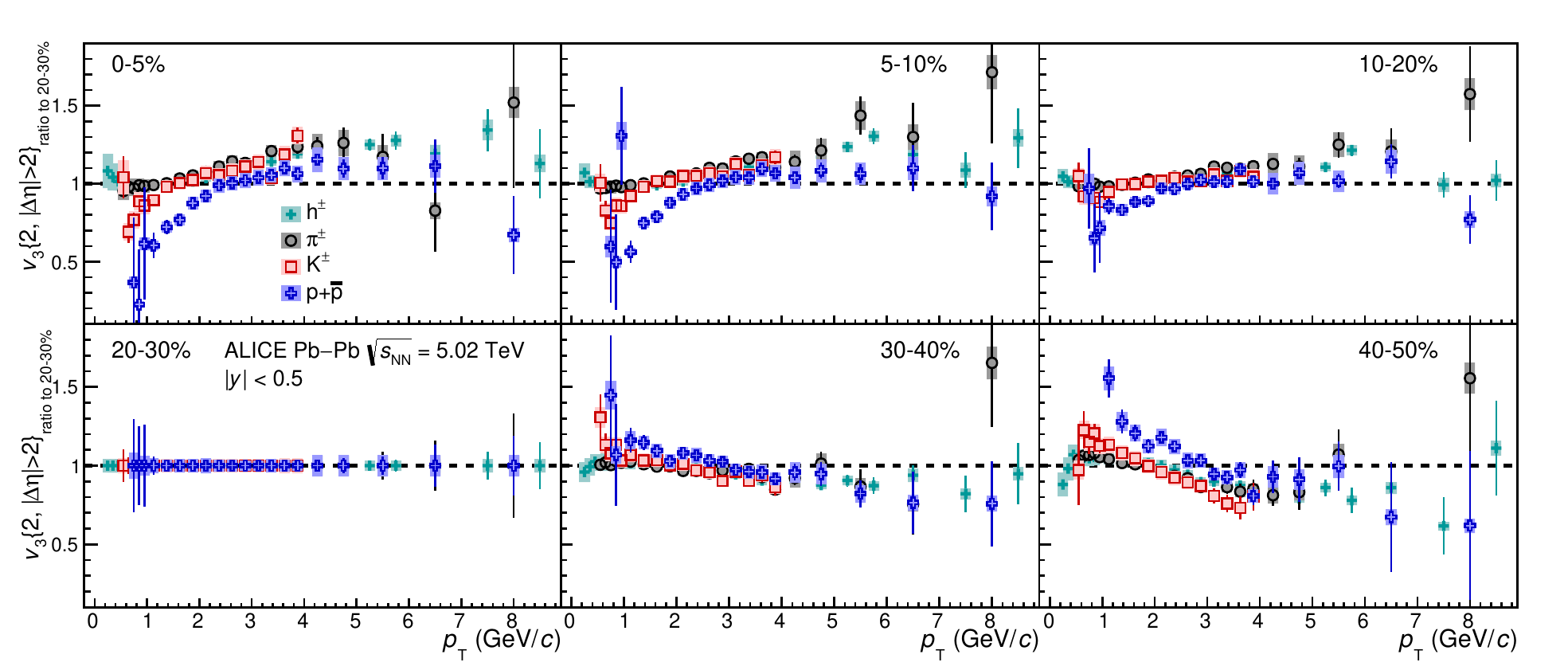}
    \caption{(Colour online) Centrality dependence of $\vthreept{}_{\rm ratio~to~20-30\%}$ for \pipm{}, \kapm{}, p+\pbar{}, and inclusive charged particles \cite{Acharya:2018lmh}. Statistical and systematic uncertainties are shown as bars and boxes, respectively.}
    \label{fig:HistV3PtPIDRat2030}
    \end{center}
\end{figure}

The shape evolution of elliptic and triangular flow is shown in Figs.~\ref{fig:HistV2PtPIDRat2030.pdf} and \ref{fig:HistV3PtPIDRat2030} for \pipm{}, 
\kapm{}, p+\pbar{}, and inclusive charged particles (the latter taken from \cite{Acharya:2018lmh}). For inclusive charged particles, variations in shape of about 10\% are 
observed for $\pt<3$ \GeVc{}, which increase to about 30\% for $\pt<6$ \GeVc{}. The shape evolution of \vtwopt{} shows different 
trends for \pipm{}, \kapm{}, and p+\pbar{}. While \pipm{} $v_2(\pt)_{\rm ratio~to~20-30\%}$ follows inclusive charged particle over the considered \pt{} range, 
the elliptic flow of p+\pbar{} (\kapm{}) varies between 20\% (10\%) to 250\% (55\%) at low \pt{} from most central to peripheral collisions. The variations are 
more pronounced for \vthreept{}, in particular for central collisions. The mass dependence found in the shape evolution of both $v_2$ and $v_3$ for 
$\pt<4$ \GeVc{} can be attributed to variations of the magnitude of radial flow and quark density, both being larger for central than peripheral 
collisions. Radial flow has a stronger effect on the \vn{} of heavier particles than that of lighter particles at low \pt{}, while the quark density influences the 
peak value of \vnpt{} in the coalescence model picture~\cite{coalescence,coalescence1,coalescence2}. For $\pt>4$~\GeVc{}, the shape evolution shows little (if any) particle type dependence. 

\begin{figure}
    \begin{center}
    \includegraphics[width=\textwidth]{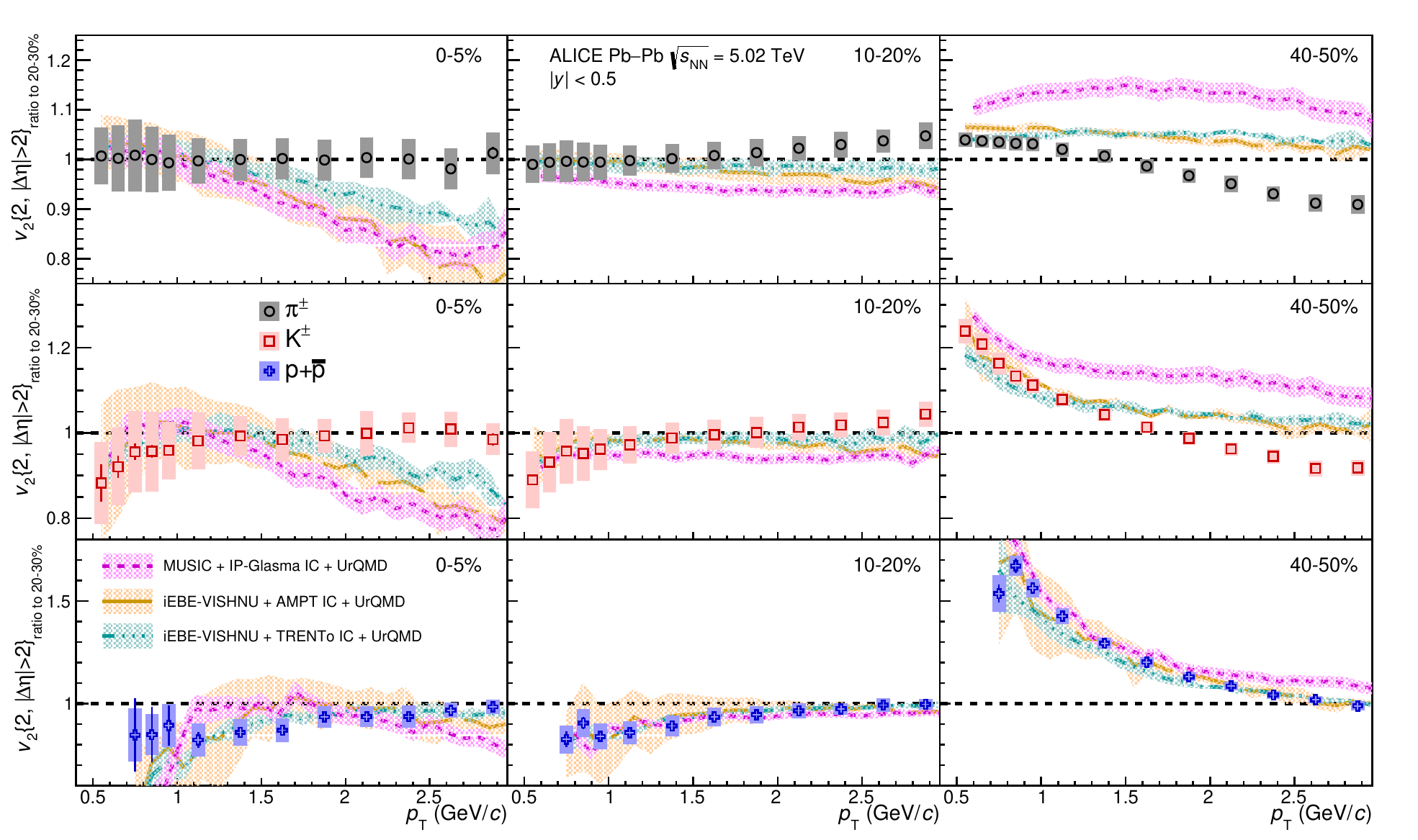}
    \caption{(Colour online) Centrality dependence of $\vtwopt{}_{\rm ratio~to~20-30\%}$ for \pipm{} (upper panels), \kapm{} (middle panels), and p+\pbar{} (lower panels) compared to hydrodynamical calculations from the MUSIC model using IP-Glasma initial conditions (magenta)~\cite{dud4}, the iEBE-VISHNU hybrid model using AMPT (orange) or \protect\trento{}~(cyan) initial conditions~\cite{wenbin}. Statistical and systematic uncertainties of the data points are shown as bars and boxes, respectively.}
    \label{fig:HistV2PtPIDCompHydBWRat2030}
    \end{center}
\end{figure}

The shape evolution of \vtwopt{} for \pipm{}, \kapm{}, and p+\pbar{} is compared to calculations from the MUSIC and iEBE-VISHNU hybrid models 
in Fig.~\ref{fig:HistV2PtPIDCompHydBWRat2030}. Both models describe the shape evolution for p+\pbar{} over the \pt{} range 
$0.7<\pt<3$~\GeVc{}. The iEBE-VISHNU model reproduces the shape evolution for \pipm{} and \kapm{} for $\pt<1.5$~\GeVc{}. Calculations from the 
MUSIC model deviate strongly from the observed shape evolution for \pipm{} and \kapm{} in peripheral collisions. 

\begin{figure}
    \begin{center}
        \includegraphics[width=.49\textwidth]{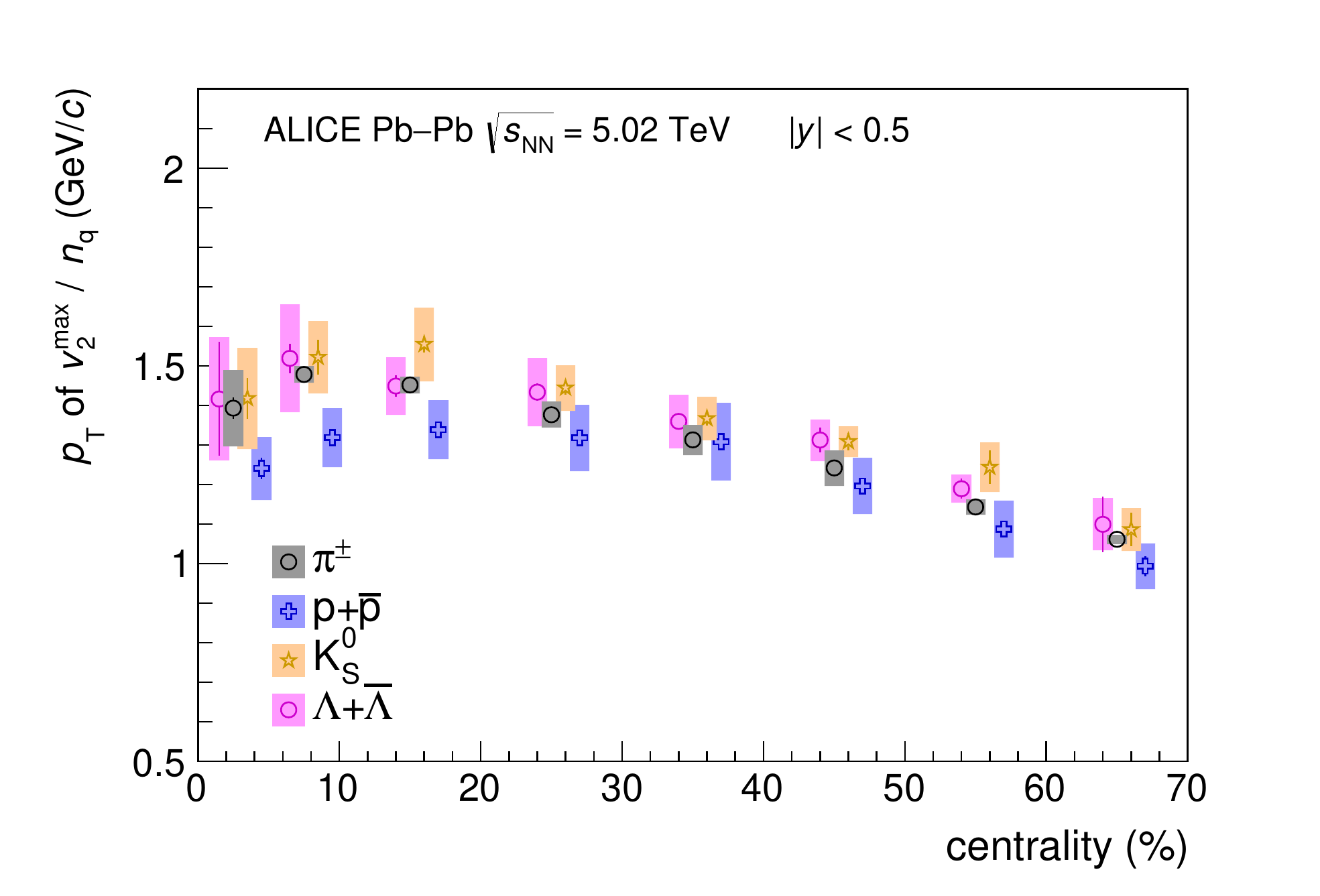}
        \includegraphics[width=.49\textwidth]{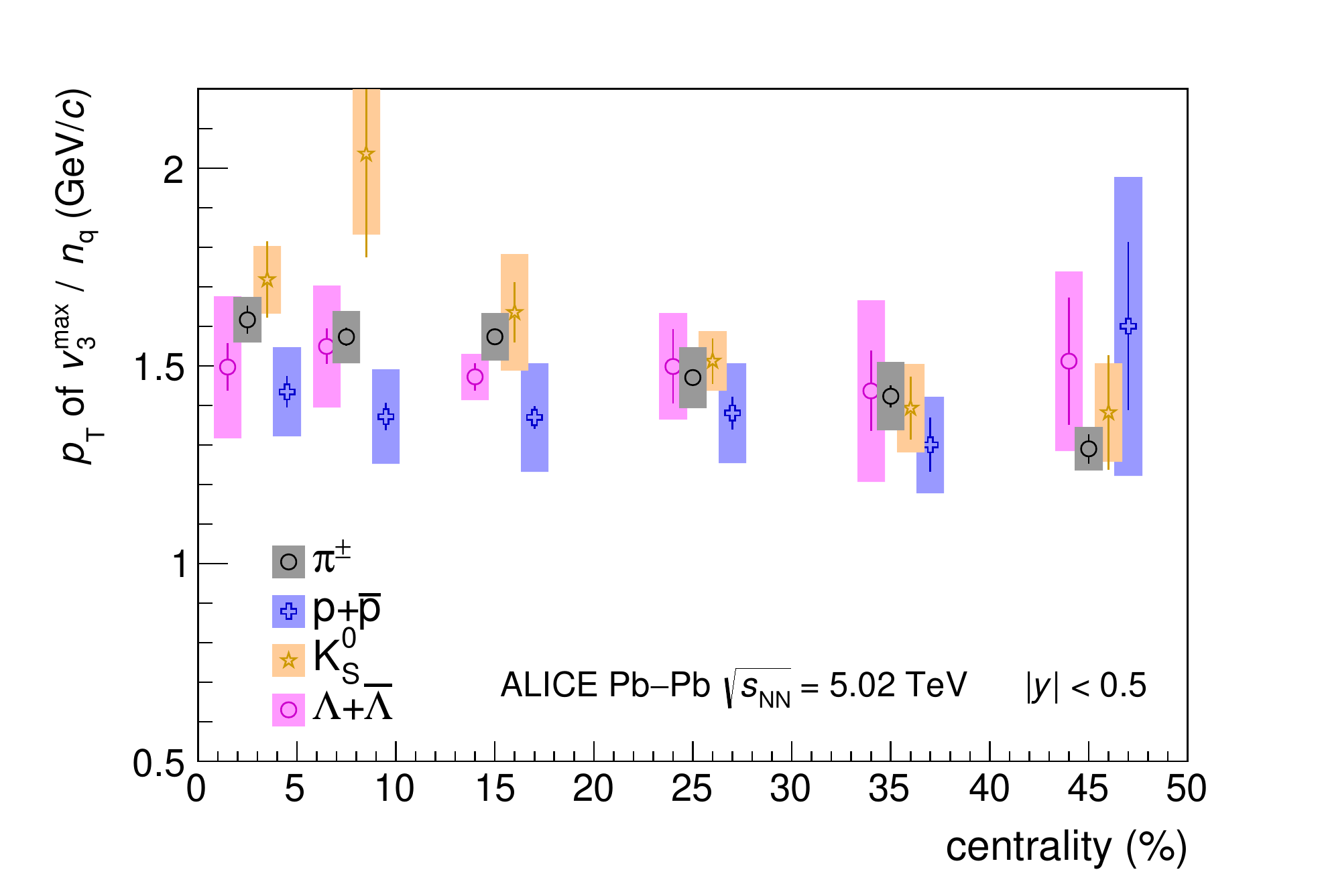}
        \caption{(Colour online) Centrality dependence of \vtwomax{} (left) and \vthreemax{} (right) divided by number of constituent quarks, $n_{\rm q}$, for \pipm{}, p+\pbar{}, \lambdas{}, and \kanull{}. Points are slightly shifted along the horizontal axis for better visibility in both panels. Statistical and systematic uncertainties are shown as bars and boxes, respectively.}
    \label{fig:snapshot23}
    \end{center}
\end{figure}

As quark density depends on centrality, the maximum \vn{} is expected to be found at higher \pt{} in more central collisions. To further quantify 
this aspect of the shape evolution of \vnpt{}, the \pt{} of \pipm{}, p+\pbar{}, \lambdas{}, and \kanull{} where \vtwopt{} and \vthreept{} reach a 
maximum, divided by number of constituent quarks $n_{\rm q}$, is reported in Fig.~\ref{fig:snapshot23} as a function of centrality. The 
$\phi$-meson and \kapm{} are not included since the kinematic range and granularity of the measurements do not allow for a reliable 
extraction of a maximum. The left panel of Fig.~\ref{fig:snapshot23} shows that the \pt{}$/n_{\rm q}$ at which \vtwopt{} reaches a 
maximum, denoted as \vtwomax{}, decreases with increasing centrality percentile for collision centralities larger than 5--10\%, following the 
expectations from the hypothesis of hadronization through coalescence. The systematic uncertainties as presented in Fig.~\ref{fig:snapshot23} 
have been evaluated directly on \vnmax{} to accurately take into account that some systematic uncertainties can be point-by-point 
correlated in \pt{}. In the 0--5\% centrality interval, there is a hint of a lower \vtwomax{} than in the 5--10\% centrality class 
for all particle species. The observed \vtwomax{} is compatible among all particle species with the exception of the p+\pbar{} \vtwomax{}, 
which is slightly lower in the 0--20\% centrality range. The right panel of Fig.~\ref{fig:snapshot23} presents \vthreemax{}, which shows, 
within the large uncertainties, a weak (if any) centrality dependence for \pipm{} and \kanull{} and no centrality dependence 
for p+\pbar{} and \lambdas{}. The \vthreemax{} is the same for the different particle species within uncertainties. 

\begin{figure}
    \begin{center}
    \includegraphics[width=\textwidth]{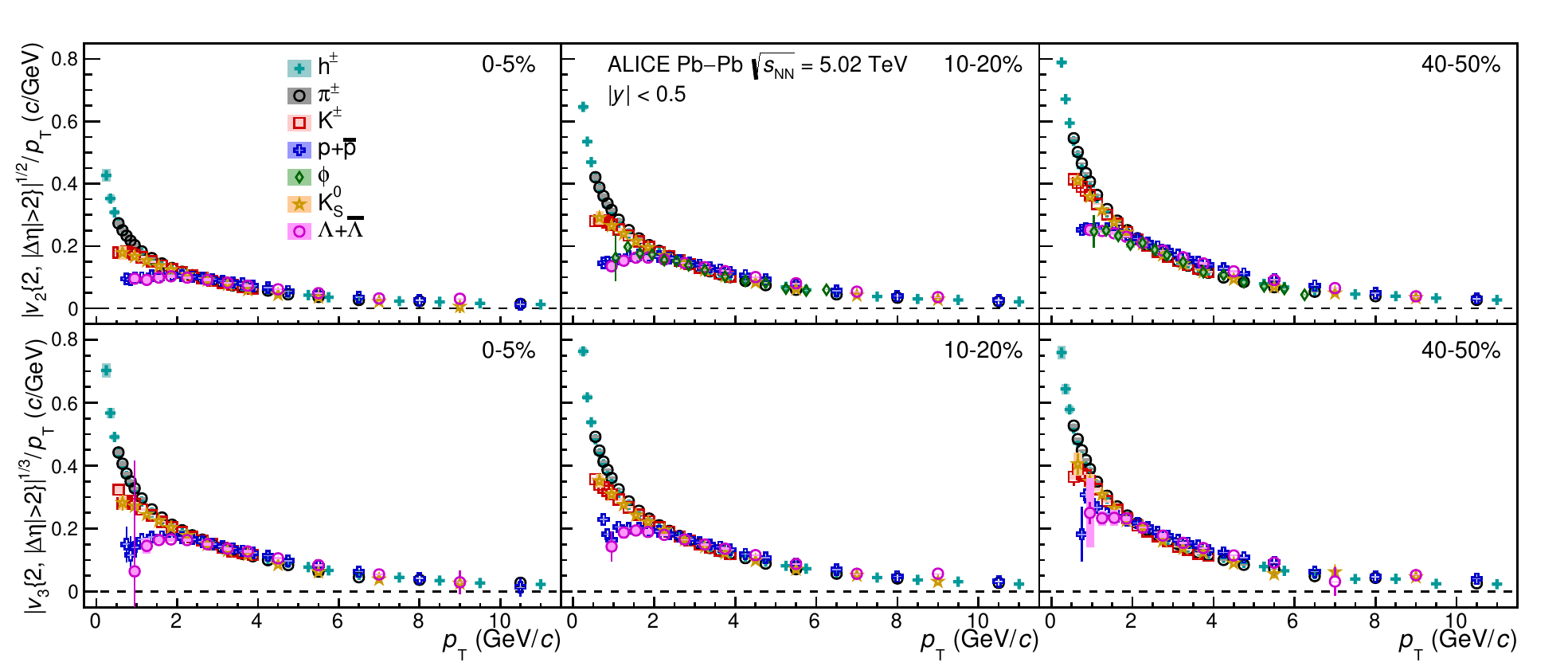}
    \caption{(Colour online) Centrality dependence of $|v_{\rm n}|^{1/{\rm n}} / \pt{}$ of inclusive charged particles \cite{Acharya:2018lmh}, \pipm{}, \kapm{}, p+\pbar{}, \lambdas{}, \kanull{}, and the $\phi$-meson for $n=2$ (upper panels) and $n=3$ (lower panels). Statistical and systematic uncertainties are shown as bars and boxes, respectively.}
    \label{fig:HistVnPtPIDRatSqRootCubicPt}
    \end{center}
\end{figure}

In the scenario of ideal hydrodynamics, \vn{} is a power law function of the radial expansion velocity of the 
medium \cite{Dinh:1999mn,Borghini:2005kd} so that $v_{\rm n} \propto \pt{}^{\rm n}$ up to $\pt{} \sim M$ for particles with 
mass $M$. Figure~\ref{fig:HistVnPtPIDRatSqRootCubicPt} shows $\vert v_{\rm n}\vert^{1/{\rm n}}/ \pt{}$ as function of \pt{} for $n = 2$ and 
$n = 3$ in various centrality intervals for inclusive charged particles \cite{Acharya:2018lmh}, \pipm{}, \kapm{}, p+\pbar{}, \lambdas{}, \kanull{}, 
and the $\phi$-meson ($n = 2$ only). When $v_{\rm n} \propto \pt{}^{\rm n}$,  the observable $\vert v_{\rm n}\vert^{1/{\rm n}}/ \pt{}$ should 
be a constant. For \pipm{} and the inclusive charged particles, the $v_{\rm n} \propto \pt{}^{\rm n}$ scaling is broken both for $v_2$ and 
$v_3$ for all centrality intervals, as is also hypothesized in \cite{Alver:2010dn}. It should be noted however that the kinematic constraints 
imposed on the measurement preclude testing the scaling hypothesis in the full relevant momentum region. The scaling holds up to 
$\pt{}~\approx~1~\GeVc{}$ for \kapm{} and \kanull{}, and up to $\pt{}~\approx~2~\GeVc{}$ for p+\pbar{}, \lambdas{}, and the 
$\phi$-meson for the 0--5\% and 10--20\% centrality intervals. Similar qualitative observations are found in the three hydrodynamical 
calculations~\cite{dud4, wenbin}.

If \vn{} indeed exhibits a power law dependence on $\pt{}^{\rm n}$,  ratios of the form of $v_{\rm n}^{1/{\rm n}}/v_{\rm m}^{1/{\rm m}}$ are 
\pt{}-independent. Previous measurements at RHIC~\cite{Adams:2003zg, Adare:2010ux} and the LHC~\cite{ATLAS:2012at,CMS:2013bza} have shown 
that the ratios $v_{\rm n}^{1/{\rm n}}/v_{\rm m}^{1/{\rm m}}$ show little to no \pt{} dependence up to about 6 \GeVc{} independent of the harmonic $n$ 
and $m$ for peripheral and semi-central collisions. However, a \pt{} dependence is observed for central collisions, which might be due to fluctuations in the initial geometry \cite{Adare:2010ux}.
\begin{figure}[t]
    \begin{center}
    \includegraphics[width=\textwidth]{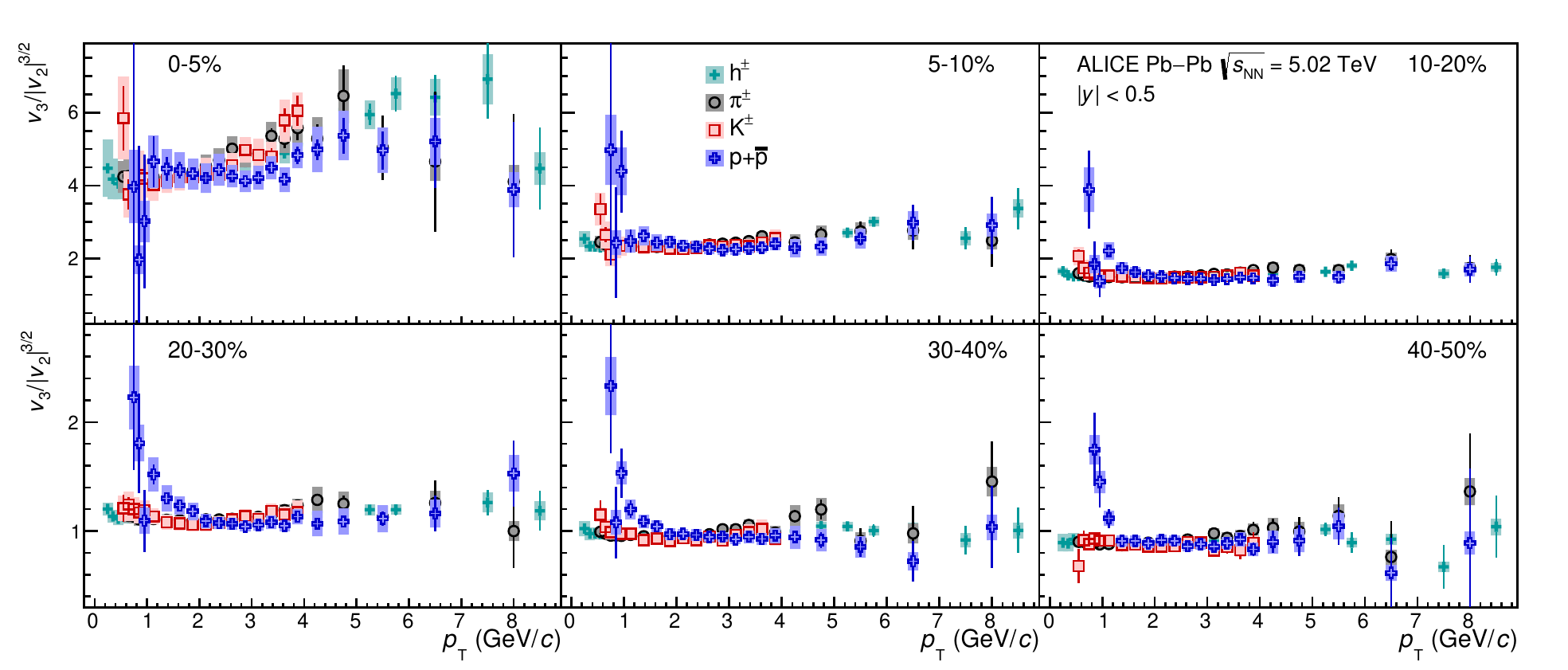}
    \caption{(Colour online) Centrality dependence of $v_3/\vert v_2\vert^{3/2}$ for inclusive charged particles \cite{Acharya:2018lmh}, \pipm{}, \kapm{}, and p+\pbar{}. Statistical and systematic uncertainties are shown as bars and boxes, respectively.}
    \label{fig:HistVnPtPIDRatV3CubicV2SqRoot}
    \end{center}
\end{figure}
\begin{figure}[t]
    \begin{center}
    \includegraphics[width=\textwidth]{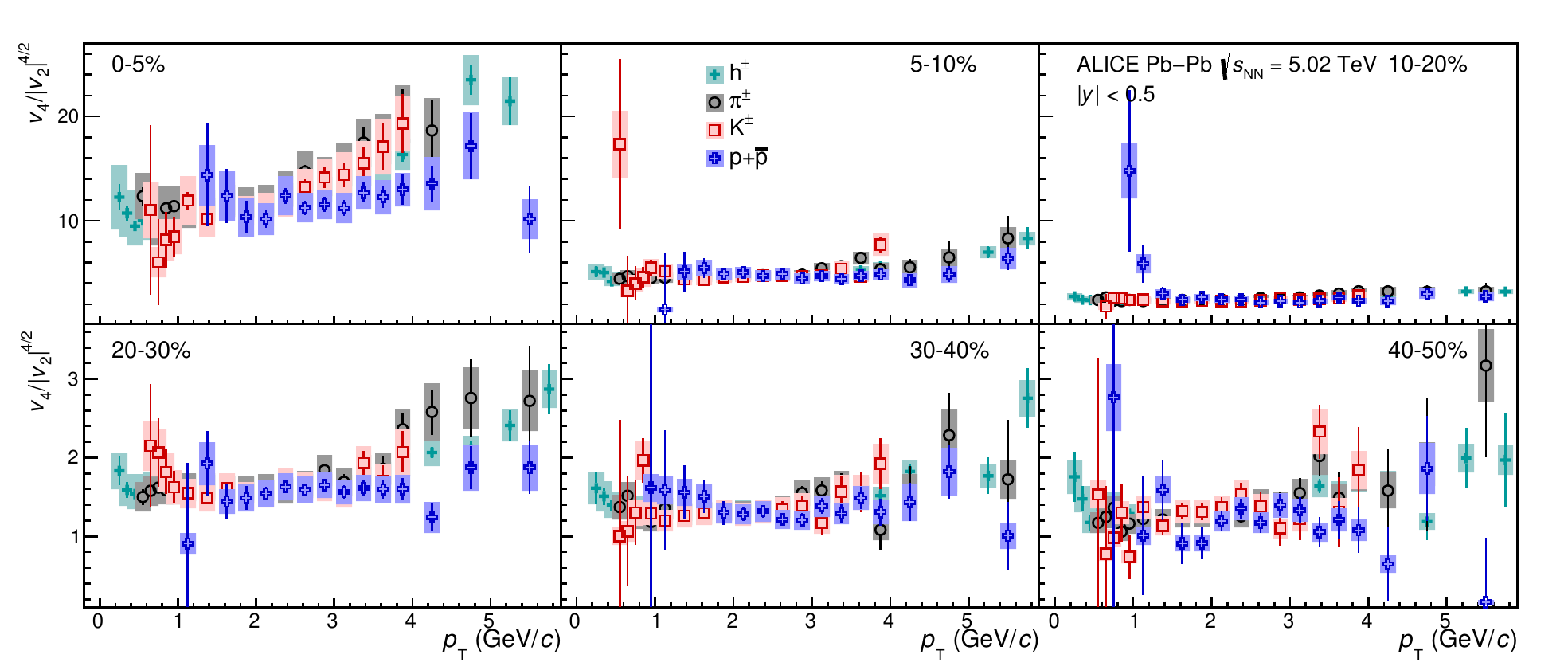}
    \caption{(Colour online) Centrality dependence of $v_4/\vert v_2\vert^{4/2}$ for inclusive charged particles \cite{Acharya:2018lmh}, \pipm{}, \kapm{}, and p+\pbar{}. Statistical and systematic uncertainties are shown as bars and boxes, respectively.}
    \label{fig:HistVnPtPIDRatV4CubicV2SqRoot}
    \end{center}
\end{figure}
\begin{figure}[t]
    \begin{center}
    \includegraphics[width=\textwidth]{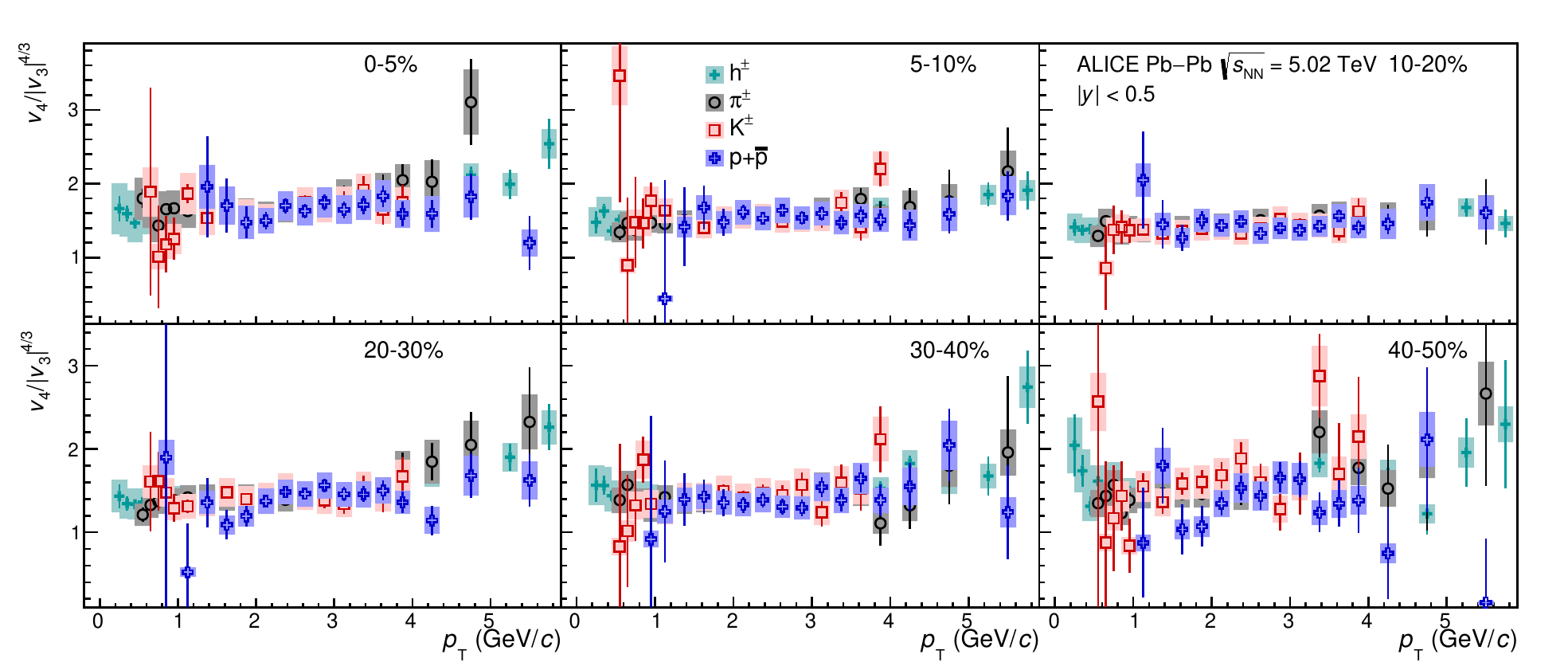}
    \caption{(Colour online) Centrality dependence of $v_4/\vert v_3\vert^{4/3}$ for inclusive charged particles \cite{Acharya:2018lmh}, \pipm{}, \kapm{}, and p+\pbar{}. Statistical and systematic uncertainties are shown as bars and boxes, respectively.}
    \label{fig:HistVnPtPIDRatV4CubicV3SqRoot}
    \end{center}
\end{figure}
The ratios $v_3/\vert v_2\vert^{3/2}$, $v_4/\vert v_2\vert^{4/2}$, and $v_4/\vert v_3\vert^{4/3}$, which probe the same scaling but are 
in practice more sensitive, are shown in Figs.~\ref{fig:HistVnPtPIDRatV3CubicV2SqRoot},~\ref{fig:HistVnPtPIDRatV4CubicV2SqRoot}, 
and \ref{fig:HistVnPtPIDRatV4CubicV3SqRoot}, respectively. For each figure, $v_{\rm n}/\vert v_{\rm m}\vert^{\rm n/m}$ is shown for 
inclusive charged particles \cite{Acharya:2018lmh}, \pipm{}, \kapm{} and p+\pbar{} in various centrality intervals. For 
$v_3/\vert v_2\vert^{3/2}$ and $v_4/\vert v_2\vert^{4/2}$, no obvious \pt{} dependence is found for inclusive charged particles 
between 5--50\% collision centrality. For the 0--5\% centrality class, the ratios are flat for $\pt<3$ \GeVc{} and rise monotonically for 
higher momenta. No particle type dependence of the ratios is found for $\pt{} >$ 1.5 \GeVc{}, below which the ratios for 
p+\pbar{} $v_{\rm n}$ rise. This rise of the p+\pbar{} $v_{\rm n}$ ratios can be attributed to an 
increase of radial flow which affects the independent harmonics differently. For the ratio $v_4/\vert v_3\vert^{4/3}$, no \pt{} dependence 
is observed over the full centrality range. Large statistical uncertainties do not allow conclusions to be drawn on the behaviour of 
p+\pbar{} $v_n$ in the $v_4/\vert v_3\vert^{4/3}$ ratio. 

\subsection{Comparison with $v_{\rm n}$ of identified particles at \sqrtSnn{} = 2.76 TeV}
\label{sec:resrun1}

\begin{figure}
     \includegraphics[width=\textwidth]{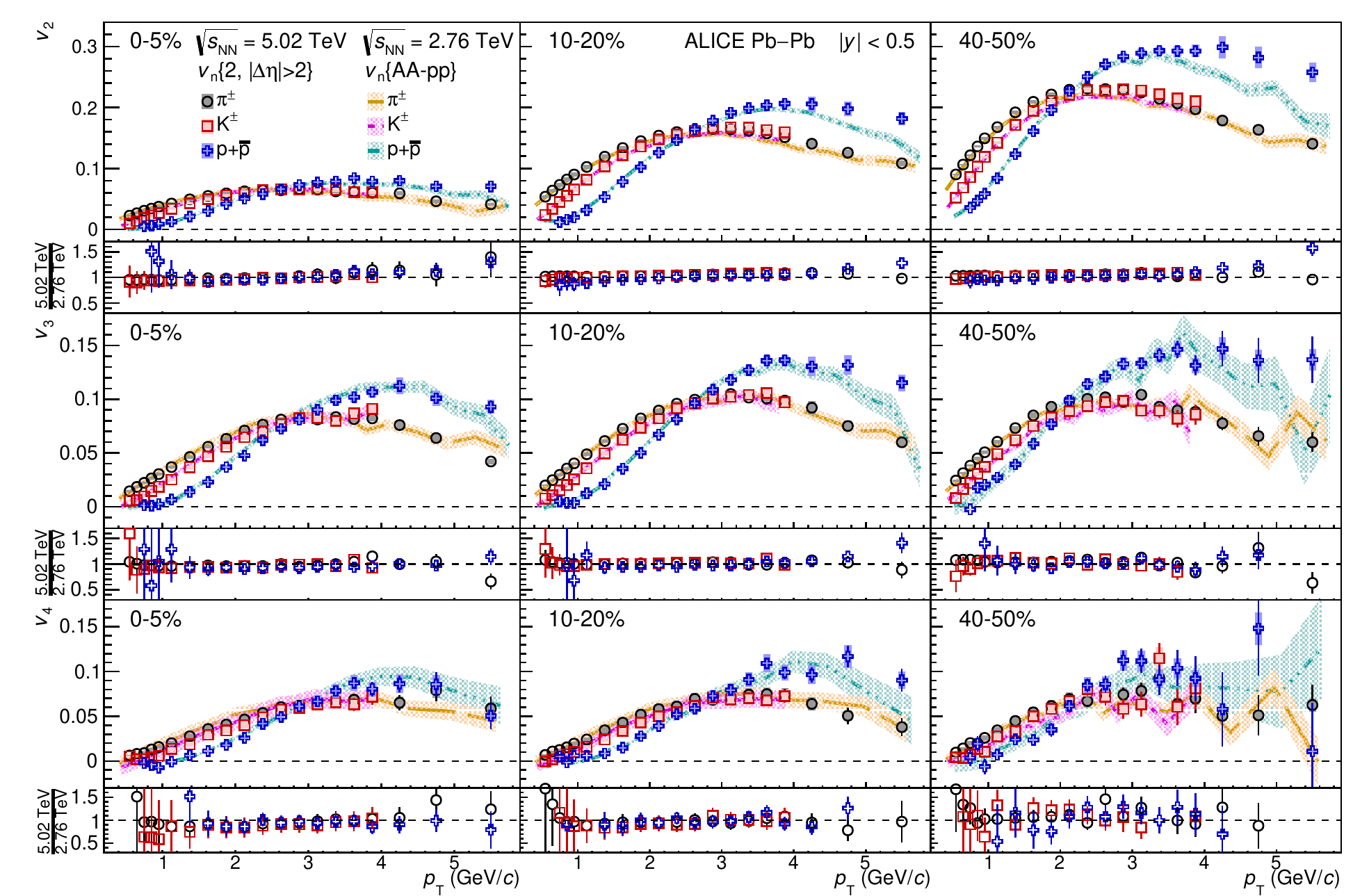}\\
     \caption{(Colour online) The $\pt$-differential $v_2$ (top), $v_3$ (middle), and $v_4$ (bottom) of \pipm{}, \kapm{}, and p+\pbar{} compared to ALICE measurements performed in Pb--Pb collisions at \sqrtSnn{} = 2.76~TeV (coloured bands)~\cite{highham} for the 0--5\%, 10--20\%, and 40--50\% centrality classes. For the measurements at \sqrtSnn{} = 5.02~TeV, statistical and systematic uncertainties are shown as bars and boxes, respectively. For the measurements at \sqrtSnn{} = 2.76~TeV, the thickness of the bands corresponds to the quadratic sum of statistical and systematic uncertainties. The ratios of measurements at \sqrtSnn{} = 5.02~TeV to a cubic spline fit to the measurements at \sqrtSnn{} = 2.76~TeV are also presented for clarity.}
     \label{fig:run14}
\end{figure}

The transport properties and initial condition models can be further constrained by studying the energy dependence of anisotropic 
flow. Figure~\ref{fig:run14} presents the \vtwopt{}, \vthreept{}, and \vfourpt{} of \pipm{}, \kapm{}, and p+\pbar{} compared to ALICE 
measurements performed at \sqrtSnn{} = 2.76 TeV~\cite{highham}. 

The $v_n$ coefficients at \sqrtSnn{} = 2.76 TeV have been measured using the scalar product method, taking the particle of 
interest under study and the charged reference particles from different, non-overlapping pseudorapidity regions between 
$\vert \eta \vert < 0.8$. Assuming no anisotropic flow in minimum bias pp collisions at the same collision energy, the non-flow 
contributions are estimated from minimum bias pp collisions and subtracted from the measured $v_{\rm n}$ coefficients. Ratios 
of the measurements presented in this paper to a cubic spline fit to the ones performed at \sqrtSnn{} = 2.76 TeV are given in the 
figure for each presented centrality interval and flow coefficient. The uncertainties in these ratios are obtained by summing the 
statistical and systematic uncertainties on the independent measurements in quadrature, and propagating the obtained uncertainties 
as uncorrelated. 

An increase of radial flow with increasing collision energy is expected to lead to a suppression of $v_{\rm n}$ at low \pt{}, an effect which 
would be most pronounced for heavier particles. Although a possible suppression of p+\pbar{} $v_{\rm n}$ at \sqrtSnn{} = 5.02 TeV can be seen 
between $1~\lesssim~\pt{}~\lesssim~3~\GeVc{}$ in central collisions and additionally for \vtwopt{} of \pipm{} and \kapm{} at the same 
centrality interval, the precision of the results does not allow for conclusions to be drawn as the measurements at different collision 
energies are compatible within uncertainties.

\begin{figure}
     \includegraphics[width=\textwidth]{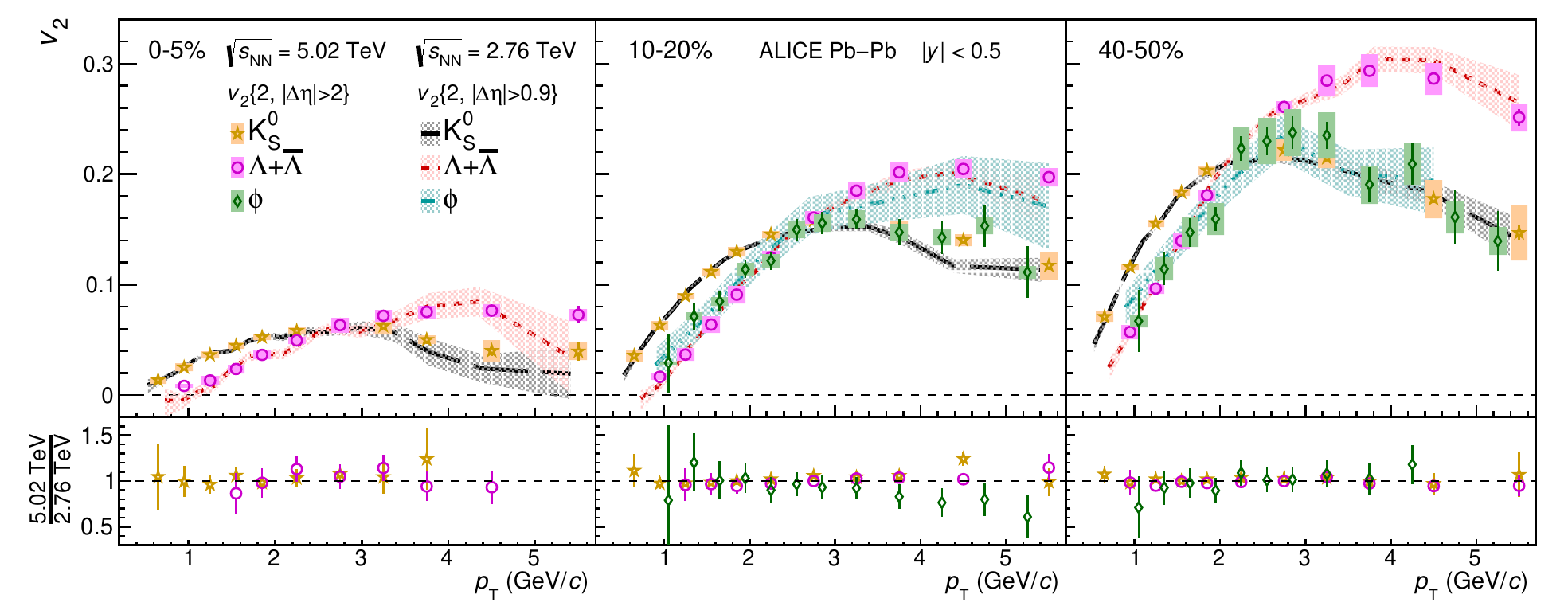}
     \caption{(Colour online) The $\pt$-differential $v_2$ of \lambdas{}, \kanull{}, and the $\phi$-meson compared to ALICE measurements performed in Pb--Pb collisions at \sqrtSnn{} = 2.76~TeV (coloured bands)~\cite{Abelev:2014pua} for the 0--5\%, 10--20\%, and 40--50\% centrality classes. For the measurements at \sqrtSnn{} = 5.02~TeV, statistical and systematic uncertainties are shown as bars and boxes, respectively. For the measurements at \sqrtSnn{} = 2.76~TeV, the thickness of the bands corresponds to the quadratic sum of statistical and systematic uncertainties. The ratios of measurements at \sqrtSnn{} = 5.02~TeV to a cubic spline fit to the measurements at \sqrtSnn{} = 2.76~TeV are also presented for clarity.}
     \label{fig:run15}
\end{figure}

Figure~\ref{fig:run15} shows the \vtwopt{} of \lambdas{}, \kanull{}, and the $\phi$-meson compared to ALICE measurements performed 
at \sqrtSnn{} = 2.76 TeV~\cite{Abelev:2014pua}, where the $v_2$ coefficients at \sqrtSnn{} = 2.76 TeV have been measured using the 
scalar product method with an $|\Delta\eta|>0.9$ gap to suppress non-flow. No differences are observed between the \kanull{} and 
\lambdas{} \vtwopt{} measured at two different collision energies. The strongly improved precision of 
the $\phi$-meson measurement at \sqrtSnn{} = 5.02 TeV, both in terms of statistical uncertainty and granularity in \pt{}, shows that the 
\vtwopt{} follows a mass ordering at low \pt{} and groups with mesons after \pt $\approx$ 3 \GeVc{} for all centrality intervals. 

\section{Summary}
\label{sec:summary}

In summary, the elliptic, triangular, and quadrangular flow coefficients of \pipm{}, \kapm{}, p+\pbar{}, \lambdas{}, \kanull{}, and the $\phi$-meson 
have been measured in Pb--Pb collisions at $\sqrt{s_{\rm NN}}=5.02$~TeV over a broad range of transverse momentum and in various centrality 
ranges. The precision of these measurements provide constraints for initial-state fluctuations and transport coefficients of the medium. The 
magnitude of $v_{\rm n}$ increases with decreasing centrality up to the 40--50\% centrality interval for 
all particle species. This increase is stronger for $v_2$ than for $v_3$ and $v_4$, which indicates that collision geometry dominates the 
generation of elliptic flow while higher flow coefficients are mainly generated by event-by-event fluctuations in the initial nucleon and gluon 
densities. This interpretation is also supported by the non-zero, positive $v_{\rm n}$ found in the 0--1\% centrality interval. In most central 
collisions (i.e.\ 0--1\% and 0--5\%), $v_3$ and $v_4$ reach a similar magnitude as $v_2$ at different $\pt$ values depending on particle 
mass, after which they increase gradually. For $\pt<3$~\GeVc{}, the $v_{\rm n}$ coefficients show a mass ordering consistent with an interplay 
between anisotropic flow and the isotropic expansion (radial flow) of the collision system. In this transverse momentum range, the 
iEBE-VISHNU hydrodynamical calculations describe the measured $v_{\rm n}$ of \pipm{}, \kapm{}, and p+\pbar{} fairly well 
for $\pt<2.5$~\GeVc{}, while MUSIC reproduces the measurements for $\pt<1$~\GeVc{}. It should be noted that neither of the presented 
hydrodynamical models is able to fully describe the measurements. At intermediate transverse 
momenta ($3<\pt<$ 8-10~\GeVc{}), particles show an approximate grouping by the number of constituent quarks at the level of 
$\pm 20\%$ for all flow coefficients in the 0--50\% centrality range. The $\phi$-meson $v_2$, which tests both particle mass 
dependence and type scaling, follows p+\pbar{} $v_2$ at low $\pt$ and \pipm{} $v_2$ at intermediate \pt{}. The baryon $v_{\rm n}$ has 
a magnitude larger than that of mesons for $\pt<$ 8-10~\GeVc{}, indicating that the particle type dependence persists up to high 
$\pt$. For $\pt>10$~\GeVc{}, the $v_2$ of p+\pbar{} is compatible with that of \pipm{} within uncertainties. The shape evolution of \vtwopt{} as 
function of centrality shows different trends for \pipm{}, \kapm{}, and p+\pbar{} and varies between 20\% (10\%) to 250\% (55\%) 
for p+\pbar{} (\kapm{}) at low \pt{} from most central to peripheral collisions; variations are more pronounced for \vthreept{}, in particular for 
central collisions. Ratios $v_3/\vert v_2\vert^{3/2}$ and $v_4/\vert v_2\vert^{4/2}$ are flat for $\pt<3$ \GeVc{} and rise monotonically 
for higher momenta for the 0-5\% centrality class. No particle type dependence of the ratios is found for $\pt{} >$ 1.5 \GeVc{}, below which 
the ratios for p+\pbar{} $v_{\rm n}$ rise, which can be attributed to an increase of radial flow which affects the independent harmonics 
differently. For the ratio $v_4/\vert v_3\vert^{4/3}$, no \pt{} dependence is observed over the full centrality range. The measurements are 
compatible with those performed in Pb--Pb collisions at \sqrtSnn{} = 2.76~TeV within uncertainties.

\newenvironment{acknowledgement}{\relax}{\relax}
\begin{acknowledgement}
\section*{Acknowledgements}

The ALICE Collaboration would like to thank all its engineers and technicians for their invaluable contributions to the construction of the experiment and the CERN accelerator teams for the outstanding performance of the LHC complex.
The ALICE Collaboration gratefully acknowledges the resources and support provided by all Grid centres and the Worldwide LHC Computing Grid (WLCG) collaboration.
The ALICE Collaboration acknowledges the following funding agencies for their support in building and running the ALICE detector:
A. I. Alikhanyan National Science Laboratory (Yerevan Physics Institute) Foundation (ANSL), State Committee of Science and World Federation of Scientists (WFS), Armenia;
Austrian Academy of Sciences and Nationalstiftung f\"{u}r Forschung, Technologie und Entwicklung, Austria;
Ministry of Communications and High Technologies, National Nuclear Research Center, Azerbaijan;
Conselho Nacional de Desenvolvimento Cient\'{\i}fico e Tecnol\'{o}gico (CNPq), Universidade Federal do Rio Grande do Sul (UFRGS), Financiadora de Estudos e Projetos (Finep) and Funda\c{c}\~{a}o de Amparo \`{a} Pesquisa do Estado de S\~{a}o Paulo (FAPESP), Brazil;
Ministry of Science \& Technology of China (MSTC), National Natural Science Foundation of China (NSFC) and Ministry of Education of China (MOEC) , China;
Ministry of Science and Education, Croatia;
Ministry of Education, Youth and Sports of the Czech Republic, Czech Republic;
The Danish Council for Independent Research | Natural Sciences, the Carlsberg Foundation and Danish National Research Foundation (DNRF), Denmark;
Helsinki Institute of Physics (HIP), Finland;
Commissariat \`{a} l'Energie Atomique (CEA) and Institut National de Physique Nucl\'{e}aire et de Physique des Particules (IN2P3) and Centre National de la Recherche Scientifique (CNRS), France;
Bundesministerium f\"{u}r Bildung, Wissenschaft, Forschung und Technologie (BMBF) and GSI Helmholtzzentrum f\"{u}r Schwerionenforschung GmbH, Germany;
General Secretariat for Research and Technology, Ministry of Education, Research and Religions, Greece;
National Research, Development and Innovation Office, Hungary;
Department of Atomic Energy Government of India (DAE), Department of Science and Technology, Government of India (DST), University Grants Commission, Government of India (UGC) and Council of Scientific and Industrial Research (CSIR), India;
Indonesian Institute of Science, Indonesia;
Centro Fermi - Museo Storico della Fisica e Centro Studi e Ricerche Enrico Fermi and Istituto Nazionale di Fisica Nucleare (INFN), Italy;
Institute for Innovative Science and Technology , Nagasaki Institute of Applied Science (IIST), Japan Society for the Promotion of Science (JSPS) KAKENHI and Japanese Ministry of Education, Culture, Sports, Science and Technology (MEXT), Japan;
Consejo Nacional de Ciencia (CONACYT) y Tecnolog\'{i}a, through Fondo de Cooperaci\'{o}n Internacional en Ciencia y Tecnolog\'{i}a (FONCICYT) and Direcci\'{o}n General de Asuntos del Personal Academico (DGAPA), Mexico;
Nederlandse Organisatie voor Wetenschappelijk Onderzoek (NWO), Netherlands;
The Research Council of Norway, Norway;
Commission on Science and Technology for Sustainable Development in the South (COMSATS), Pakistan;
Pontificia Universidad Cat\'{o}lica del Per\'{u}, Peru;
Ministry of Science and Higher Education and National Science Centre, Poland;
Korea Institute of Science and Technology Information and National Research Foundation of Korea (NRF), Republic of Korea;
Ministry of Education and Scientific Research, Institute of Atomic Physics and Romanian National Agency for Science, Technology and Innovation, Romania;
Joint Institute for Nuclear Research (JINR), Ministry of Education and Science of the Russian Federation and National Research Centre Kurchatov Institute, Russia;
Ministry of Education, Science, Research and Sport of the Slovak Republic, Slovakia;
National Research Foundation of South Africa, South Africa;
Centro de Aplicaciones Tecnol\'{o}gicas y Desarrollo Nuclear (CEADEN), Cubaenerg\'{\i}a, Cuba and Centro de Investigaciones Energ\'{e}ticas, Medioambientales y Tecnol\'{o}gicas (CIEMAT), Spain;
Swedish Research Council (VR) and Knut \& Alice Wallenberg Foundation (KAW), Sweden;
European Organization for Nuclear Research, Switzerland;
National Science and Technology Development Agency (NSDTA), Suranaree University of Technology (SUT) and Office of the Higher Education Commission under NRU project of Thailand, Thailand;
Turkish Atomic Energy Agency (TAEK), Turkey;
National Academy of  Sciences of Ukraine, Ukraine;
Science and Technology Facilities Council (STFC), United Kingdom;
National Science Foundation of the United States of America (NSF) and United States Department of Energy, Office of Nuclear Physics (DOE NP), United States of America.
\end{acknowledgement}

\bibliographystyle{utphys}
\bibliography{references}

\newpage
\appendix
\section{The ALICE Collaboration}
\label{app:collab}

\begingroup
\small
\begin{flushleft}
S.~Acharya\Irefn{org139}\And 
F.T.-.~Acosta\Irefn{org22}\And 
D.~Adamov\'{a}\Irefn{org94}\And 
J.~Adolfsson\Irefn{org81}\And 
M.M.~Aggarwal\Irefn{org98}\And 
G.~Aglieri Rinella\Irefn{org36}\And 
M.~Agnello\Irefn{org33}\And 
N.~Agrawal\Irefn{org49}\And 
Z.~Ahammed\Irefn{org139}\And 
S.U.~Ahn\Irefn{org77}\And 
S.~Aiola\Irefn{org144}
% ...
\And
\renewcommand\labelenumi{\textsuperscript{\theenumi}~}

\section*{Affiliation notes}
\renewcommand\theenumi{\roman{enumi}}
\begin{Authlist}
\item \Adef{org*}Deceased
\item \Adef{orgI}Dipartimento DET del Politecnico di Torino, Turin, Italy
\item \Adef{orgII}M.V. Lomonosov Moscow State University, D.V. Skobeltsyn Institute of Nuclear, Physics, Moscow, Russia
\item \Adef{orgIII}Department of Applied Physics, Aligarh Muslim University, Aligarh, India
\item \Adef{orgIV}Institute of Theoretical Physics, University of Wroclaw, Poland
\end{Authlist}

\section*{Collaboration Institutes}
\renewcommand\theenumi{\arabic{enumi}~}
\begin{Authlist}
\item \Idef{org1}A.I. Alikhanyan National Science Laboratory (Yerevan Physics Institute) Foundation, Yerevan, Armenia
\item \Idef{org2}Benem\'{e}rita Universidad Aut\'{o}noma de Puebla, Puebla, Mexico
\item \Idef{org3}Bogolyubov Institute for Theoretical Physics, National Academy of Sciences of Ukraine, Kiev, Ukraine
\item \Idef{org4}Bose Institute, Department of Physics  and Centre for Astroparticle Physics and Space Science (CAPSS), Kolkata, India
\item \Idef{org5}Budker Institute for Nuclear Physics, Novosibirsk, Russia
\item \Idef{org6}California Polytechnic State University, San Luis Obispo, California, United States
\item \Idef{org7}Central China Normal University, Wuhan, China
\item \Idef{org8}Centre de Calcul de l'IN2P3, Villeurbanne, Lyon, France
\item \Idef{org9}Centro de Aplicaciones Tecnol\'{o}gicas y Desarrollo Nuclear (CEADEN), Havana, Cuba
\item \Idef{org10}Centro de Investigaci\'{o}n y de Estudios Avanzados (CINVESTAV), Mexico City and M\'{e}rida, Mexico
\item \Idef{org11}Centro Fermi - Museo Storico della Fisica e Centro Studi e Ricerche ``Enrico Fermi', Rome, Italy
\item \Idef{org12}Chicago State University, Chicago, Illinois, United States
\item \Idef{org13}China Institute of Atomic Energy, Beijing, China
\item \Idef{org14}Chonbuk National University, Jeonju, Republic of Korea
\item \Idef{org15}Comenius University Bratislava, Faculty of Mathematics, Physics and Informatics, Bratislava, Slovakia
\item \Idef{org16}COMSATS Institute of Information Technology (CIIT), Islamabad, Pakistan
\item \Idef{org17}Creighton University, Omaha, Nebraska, United States
\item \Idef{org18}Department of Physics, Aligarh Muslim University, Aligarh, India
\item \Idef{org19}Department of Physics, Ohio State University, Columbus, Ohio, United States
\item \Idef{org20}Department of Physics, Pusan National University, Pusan, Republic of Korea
\item \Idef{org21}Department of Physics, Sejong University, Seoul, Republic of Korea
\item \Idef{org22}Department of Physics, University of California, Berkeley, California, United States
\item \Idef{org23}Department of Physics, University of Oslo, Oslo, Norway
\item \Idef{org24}Department of Physics and Technology, University of Bergen, Bergen, Norway
\item \Idef{org25}Dipartimento di Fisica dell'Universit\`{a} 'La Sapienza' and Sezione INFN, Rome, Italy
\item \Idef{org26}Dipartimento di Fisica dell'Universit\`{a} and Sezione INFN, Cagliari, Italy
\item \Idef{org27}Dipartimento di Fisica dell'Universit\`{a} and Sezione INFN, Trieste, Italy
\item \Idef{org28}Dipartimento di Fisica dell'Universit\`{a} and Sezione INFN, Turin, Italy
\item \Idef{org29}Dipartimento di Fisica e Astronomia dell'Universit\`{a} and Sezione INFN, Bologna, Italy
\item \Idef{org30}Dipartimento di Fisica e Astronomia dell'Universit\`{a} and Sezione INFN, Catania, Italy
\item \Idef{org31}Dipartimento di Fisica e Astronomia dell'Universit\`{a} and Sezione INFN, Padova, Italy
\item \Idef{org32}Dipartimento di Fisica `E.R.~Caianiello' dell'Universit\`{a} and Gruppo Collegato INFN, Salerno, Italy
\item \Idef{org33}Dipartimento DISAT del Politecnico and Sezione INFN, Turin, Italy
\item \Idef{org34}Dipartimento di Scienze e Innovazione Tecnologica dell'Universit\`{a} del Piemonte Orientale and INFN Sezione di Torino, Alessandria, Italy
\item \Idef{org35}Dipartimento Interateneo di Fisica `M.~Merlin' and Sezione INFN, Bari, Italy
\item \Idef{org36}European Organization for Nuclear Research (CERN), Geneva, Switzerland
\item \Idef{org37}Faculty of Electrical Engineering, Mechanical Engineering and Naval Architecture, University of Split, Split, Croatia
\item \Idef{org38}Faculty of Engineering and Science, Western Norway University of Applied Sciences, Bergen, Norway
\item \Idef{org39}Faculty of Nuclear Sciences and Physical Engineering, Czech Technical University in Prague, Prague, Czech Republic
\item \Idef{org40}Faculty of Science, P.J.~\v{S}af\'{a}rik University, Ko\v{s}ice, Slovakia
\item \Idef{org41}Frankfurt Institute for Advanced Studies, Johann Wolfgang Goethe-Universit\"{a}t Frankfurt, Frankfurt, Germany
\item \Idef{org42}Gangneung-Wonju National University, Gangneung, Republic of Korea
\item \Idef{org43}Gauhati University, Department of Physics, Guwahati, India
\item \Idef{org44}Helmholtz-Institut f\"{u}r Strahlen- und Kernphysik, Rheinische Friedrich-Wilhelms-Universit\"{a}t Bonn, Bonn, Germany
\item \Idef{org45}Helsinki Institute of Physics (HIP), Helsinki, Finland
\item \Idef{org46}Hiroshima University, Hiroshima, Japan
\item \Idef{org47}Hochschule Worms, Zentrum  f\"{u}r Technologietransfer und Telekommunikation (ZTT), Worms, Germany
\item \Idef{org48}Horia Hulubei National Institute of Physics and Nuclear Engineering, Bucharest, Romania
\item \Idef{org49}Indian Institute of Technology Bombay (IIT), Mumbai, India
\item \Idef{org50}Indian Institute of Technology Indore, Indore, India
\item \Idef{org51}Indonesian Institute of Sciences, Jakarta, Indonesia
\item \Idef{org52}INFN, Laboratori Nazionali di Frascati, Frascati, Italy
\item \Idef{org53}INFN, Sezione di Bari, Bari, Italy
\item \Idef{org54}INFN, Sezione di Bologna, Bologna, Italy
\item \Idef{org55}INFN, Sezione di Cagliari, Cagliari, Italy
\item \Idef{org56}INFN, Sezione di Catania, Catania, Italy
\item \Idef{org57}INFN, Sezione di Padova, Padova, Italy
\item \Idef{org58}INFN, Sezione di Roma, Rome, Italy
\item \Idef{org59}INFN, Sezione di Torino, Turin, Italy
\item \Idef{org60}INFN, Sezione di Trieste, Trieste, Italy
\item \Idef{org61}Inha University, Incheon, Republic of Korea
\item \Idef{org62}Institut de Physique Nucl\'{e}aire d'Orsay (IPNO), Institut National de Physique Nucl\'{e}aire et de Physique des Particules (IN2P3/CNRS), Universit\'{e} de Paris-Sud, Universit\'{e} Paris-Saclay, Orsay, France
\item \Idef{org63}Institute for Nuclear Research, Academy of Sciences, Moscow, Russia
\item \Idef{org64}Institute for Subatomic Physics, Utrecht University/Nikhef, Utrecht, Netherlands
\item \Idef{org65}Institute for Theoretical and Experimental Physics, Moscow, Russia
\item \Idef{org66}Institute of Experimental Physics, Slovak Academy of Sciences, Ko\v{s}ice, Slovakia
\item \Idef{org67}Institute of Physics, Bhubaneswar, India
\item \Idef{org68}Institute of Physics of the Czech Academy of Sciences, Prague, Czech Republic
\item \Idef{org69}Institute of Space Science (ISS), Bucharest, Romania
\item \Idef{org70}Institut f\"{u}r Kernphysik, Johann Wolfgang Goethe-Universit\"{a}t Frankfurt, Frankfurt, Germany
\item \Idef{org71}Instituto de Ciencias Nucleares, Universidad Nacional Aut\'{o}noma de M\'{e}xico, Mexico City, Mexico
\item \Idef{org72}Instituto de F\'{i}sica, Universidade Federal do Rio Grande do Sul (UFRGS), Porto Alegre, Brazil
\item \Idef{org73}Instituto de F\'{\i}sica, Universidad Nacional Aut\'{o}noma de M\'{e}xico, Mexico City, Mexico
\item \Idef{org74}iThemba LABS, National Research Foundation, Somerset West, South Africa
\item \Idef{org75}Johann-Wolfgang-Goethe Universit\"{a}t Frankfurt Institut f\"{u}r Informatik, Fachbereich Informatik und Mathematik, Frankfurt, Germany
\item \Idef{org76}Joint Institute for Nuclear Research (JINR), Dubna, Russia
\item \Idef{org77}Korea Institute of Science and Technology Information, Daejeon, Republic of Korea
\item \Idef{org78}KTO Karatay University, Konya, Turkey
\item \Idef{org79}Laboratoire de Physique Subatomique et de Cosmologie, Universit\'{e} Grenoble-Alpes, CNRS-IN2P3, Grenoble, France
\item \Idef{org80}Lawrence Berkeley National Laboratory, Berkeley, California, United States
\item \Idef{org81}Lund University Department of Physics, Division of Particle Physics, Lund, Sweden
\item \Idef{org82}Nagasaki Institute of Applied Science, Nagasaki, Japan
\item \Idef{org83}Nara Women{'}s University (NWU), Nara, Japan
\item \Idef{org84}National and Kapodistrian University of Athens, School of Science, Department of Physics , Athens, Greece
\item \Idef{org85}National Centre for Nuclear Research, Warsaw, Poland
\item \Idef{org86}National Institute of Science Education and Research, HBNI, Jatni, India
\item \Idef{org87}National Nuclear Research Center, Baku, Azerbaijan
\item \Idef{org88}National Research Centre Kurchatov Institute, Moscow, Russia
\item \Idef{org89}Niels Bohr Institute, University of Copenhagen, Copenhagen, Denmark
\item \Idef{org90}Nikhef, National institute for subatomic physics, Amsterdam, Netherlands
\item \Idef{org91}NRC Kurchatov Institute IHEP, Protvino, Russia
\item \Idef{org92}NRNU Moscow Engineering Physics Institute, Moscow, Russia
\item \Idef{org93}Nuclear Physics Group, STFC Daresbury Laboratory, Daresbury, United Kingdom
\item \Idef{org94}Nuclear Physics Institute of the Czech Academy of Sciences, \v{R}e\v{z} u Prahy, Czech Republic
\item \Idef{org95}Oak Ridge National Laboratory, Oak Ridge, Tennessee, United States
\item \Idef{org96}Petersburg Nuclear Physics Institute, Gatchina, Russia
\item \Idef{org97}Physics department, Faculty of science, University of Zagreb, Zagreb, Croatia
\item \Idef{org98}Physics Department, Panjab University, Chandigarh, India
\item \Idef{org99}Physics Department, University of Jammu, Jammu, India
\item \Idef{org100}Physics Department, University of Rajasthan, Jaipur, India
\item \Idef{org101}Physikalisches Institut, Eberhard-Karls-Universit\"{a}t T\"{u}bingen, T\"{u}bingen, Germany
\item \Idef{org102}Physikalisches Institut, Ruprecht-Karls-Universit\"{a}t Heidelberg, Heidelberg, Germany
\item \Idef{org103}Physik Department, Technische Universit\"{a}t M\"{u}nchen, Munich, Germany
\item \Idef{org104}Research Division and ExtreMe Matter Institute EMMI, GSI Helmholtzzentrum f\"ur Schwerionenforschung GmbH, Darmstadt, Germany
\item \Idef{org105}Rudjer Bo\v{s}kovi\'{c} Institute, Zagreb, Croatia
\item \Idef{org106}Russian Federal Nuclear Center (VNIIEF), Sarov, Russia
\item \Idef{org107}Saha Institute of Nuclear Physics, Kolkata, India
\item \Idef{org108}School of Physics and Astronomy, University of Birmingham, Birmingham, United Kingdom
\item \Idef{org109}Secci\'{o}n F\'{\i}sica, Departamento de Ciencias, Pontificia Universidad Cat\'{o}lica del Per\'{u}, Lima, Peru
\item \Idef{org110}Shanghai Institute of Applied Physics, Shanghai, China
\item \Idef{org111}Stefan Meyer Institut f\"{u}r Subatomare Physik (SMI), Vienna, Austria
\item \Idef{org112}SUBATECH, IMT Atlantique, Universit\'{e} de Nantes, CNRS-IN2P3, Nantes, France
\item \Idef{org113}Suranaree University of Technology, Nakhon Ratchasima, Thailand
\item \Idef{org114}Technical University of Ko\v{s}ice, Ko\v{s}ice, Slovakia
\item \Idef{org115}Technische Universit\"{a}t M\"{u}nchen, Excellence Cluster 'Universe', Munich, Germany
\item \Idef{org116}The Henryk Niewodniczanski Institute of Nuclear Physics, Polish Academy of Sciences, Cracow, Poland
\item \Idef{org117}The University of Texas at Austin, Austin, Texas, United States
\item \Idef{org118}Universidad Aut\'{o}noma de Sinaloa, Culiac\'{a}n, Mexico
\item \Idef{org119}Universidade de S\~{a}o Paulo (USP), S\~{a}o Paulo, Brazil
\item \Idef{org120}Universidade Estadual de Campinas (UNICAMP), Campinas, Brazil
\item \Idef{org121}Universidade Federal do ABC, Santo Andre, Brazil
\item \Idef{org122}University College of Southeast Norway, Tonsberg, Norway
\item \Idef{org123}University of Cape Town, Cape Town, South Africa
\item \Idef{org124}University of Houston, Houston, Texas, United States
\item \Idef{org125}University of Jyv\"{a}skyl\"{a}, Jyv\"{a}skyl\"{a}, Finland
\item \Idef{org126}University of Liverpool, Department of Physics Oliver Lodge Laboratory , Liverpool, United Kingdom
\item \Idef{org127}University of Tennessee, Knoxville, Tennessee, United States
\item \Idef{org128}University of the Witwatersrand, Johannesburg, South Africa
\item \Idef{org129}University of Tokyo, Tokyo, Japan
\item \Idef{org130}University of Tsukuba, Tsukuba, Japan
\item \Idef{org131}Universit\'{e} Clermont Auvergne, CNRS/IN2P3, LPC, Clermont-Ferrand, France
\item \Idef{org132}Universit\'{e} de Lyon, Universit\'{e} Lyon 1, CNRS/IN2P3, IPN-Lyon, Villeurbanne, Lyon, France
\item \Idef{org133}Universit\'{e} de Strasbourg, CNRS, IPHC UMR 7178, F-67000 Strasbourg, France, Strasbourg, France
\item \Idef{org134} Universit\'{e} Paris-Saclay Centre d¿\'Etudes de Saclay (CEA), IRFU, Department de Physique Nucl\'{e}aire (DPhN), Saclay, France
\item \Idef{org135}Universit\`{a} degli Studi di Foggia, Foggia, Italy
\item \Idef{org136}Universit\`{a} degli Studi di Pavia, Pavia, Italy
\item \Idef{org137}Universit\`{a} di Brescia, Brescia, Italy
\item \Idef{org138}V.~Fock Institute for Physics, St. Petersburg State University, St. Petersburg, Russia
\item \Idef{org139}Variable Energy Cyclotron Centre, Kolkata, India
\item \Idef{org140}Warsaw University of Technology, Warsaw, Poland
\item \Idef{org141}Wayne State University, Detroit, Michigan, United States
\item \Idef{org142}Westf\"{a}lische Wilhelms-Universit\"{a}t M\"{u}nster, Institut f\"{u}r Kernphysik, M\"{u}nster, Germany
\item \Idef{org143}Wigner Research Centre for Physics, Hungarian Academy of Sciences, Budapest, Hungary
\item \Idef{org144}Yale University, New Haven, Connecticut, United States
\item \Idef{org145}Yonsei University, Seoul, Republic of Korea
\end{Authlist}
\endgroup

\end{document}